\documentclass[twocolumn,showpacs,preprintnumbers,superscriptaddress,amsmath,amssymb]{revtex4-1}

\usepackage{epsfig}
\usepackage{graphicx}
\usepackage{dcolumn}
\usepackage{bm}
\usepackage{times}
\usepackage{xcolor}
\usepackage{booktabs}
\emergencystretch=\maxdimen
\begin{document}


\title{Optimizing spreading dynamics in interconnected networks}

\author{Liming Pan}
\affiliation{Web Sciences Center, School of Computer Science and Engineering, University of Electronic Science and Technology of China, Chengdu, 611731, China}

\author{Wei Wang}
\email{wwzqbx@hotmail.com}
\affiliation{Cybersecurity Research Institute, Sichuan University, Chengdu 610065, China}
\affiliation{Web Sciences Center, School of Computer Science and Engineering, University of Electronic Science and Technology of China, Chengdu,
611731, China}

\author{Shimin Cai}
\affiliation{Web Sciences Center, School of Computer Science and Engineering, University of Electronic Science and Technology of China, Chengdu,
611731, China}

\author{Tao Zhou}
\affiliation{Web Sciences Center, School of Computer Science and Engineering, University of Electronic Science and Technology of China, Chengdu,
611731, China}

\date{\today}

\begin{abstract}
Adding edges between layers of interconnected networks is an important way to optimize the spreading dynamics. While previous studies mostly focus on the case of adding a single edge, the theoretical optimal strategy for adding multiple edges still need to be studied. In this study, based on the susceptible-infected-susceptible (SIS) model, we investigate the problem of maximizing the stationary spreading prevalence in interconnected networks. For two isolated networks, we maximize the spreading prevalence near the critical point by choosing multiple interconnecting edges. We present a theoretical analysis based on the discrete-time Markov chain approach to derive the approximate optimal strategy. The optimal inter-layer structure predicted by the strategy maximizes the spreading prevalence, meanwhile minimizes the spreading outbreak threshold for the interconnected network simultaneously. Numerical simulations on synthetic and real-world networks show that near the critical point, the proposed strategy gives better performance than connecting large degree nodes and randomly connecting.
\end{abstract}

\maketitle

\textbf{Spreading dynamics in interconnected networks relay on the inter-layer structure apart from the structure within each layer. For two given networks, how to design the interconnecting structure to optimize the spreading dynamics is a very appealing topic. Previous studies obtained the optimal strategy when considering adding a single edge in two-layer interconnected networks, while the optimal strategy of adding multiple edges lacks theoretical studies. Therefore in this study, a novel strategy is proposed to promote the spreading dynamics by adding multiple interconnecting edges for two isolated networks. Near the critical point, the spreading prevalence can be written in terms of the leading eigenvalue and the corresponding eigenvector of the adjacency matrix. Basing on an approximation scheme for the leading eigenvalue and eigenvector of the interconnected network, we optimize the spreading prevalence among all candidate edges. The optimal inter-layer structure is achieved by selecting edge candidates that are top-ranked by the product of eigenvector centrality of nodes in its two ends. Meanwhile, the optimal strategy also minimizes the outbreak threshold of the interconnected network. Numerical simulations on three pairs of synthetic networks and two pairs of real-world networks show that the strategy gives better performance than the heuristic strategies of connecting large degree nodes and of randomly connecting, especially near the critical point.}

\section{Introduction} \label{sec:intro}
In real-world social systems, individuals might communicate with others via multiple possible channels (such as Twitter, Facebook, LinkedIn). The communication relations with a specific channel can be represented by a network, where the nodes correspond to individuals, and the edges correspond to the communication relations. Therefore all the communication relations combined can be described by a multilayer network~\cite{boccaletti2014structure,gao2012networks,
kivela2014multilayer,de2016physics}, where each layer corresponds to one of the communications channels. Multilayer networks and the dynamics on them have attracted attentions from diverse areas. It has been observed that multilayer networks display distinct collective behaviors compared to that of single-layer networks~\cite{cai2018multiplex,sole2016congestion,
buono2015immunization,buldyrev2010catastrophic,
cozzo2013contact,kim2013coevolution,
lee2012correlated,wang2015evolutionary,
zhang2015explosive}. As an example, percolation processes on multilayer networks display first-order phase transitions~\cite{buldyrev2010catastrophic,valdez2013triple,
di2016recovery,di2016cascading,valdez2016failure}, which is intrinsically different from the second-order phase transitions on single-layer networks~\cite{cohen2002percolation,
serrano2006percolation,goltsev2008percolation}. For evolutionary games, multilayer networks promote cooperation better than single-layer networks~\cite{perc2017statistical,wang2013interdependent,
wang2015evolutionary}. For synchronization processes, explosive synchronization and hysteresis loop are observed on multilayer networks~\cite{zhang2015explosive}.

Spreading dynamics on multilayer networks have attracted considerable attention in recent studies~\cite{de2016physics,
wang2019coevolution,brummitt2012multiplexity,lee2014threshold,
Yagan2012,wang2018social,wang2018social2,shu2018social,
chen2018optimal}. Saumell-Mendiola, Serrano and Bogun{\'a}~\cite{saumell2012epidemic} have studied susceptible-infected-susceptible (SIS) model~\cite{pastor2001epidemic} on multilayer networks. They found that adding a small fraction of edges between layers can lead to the outbreak of epidemics while without these edges the epidemic extinct. The epidemic threshold depends on the structure of multilayer networks~\cite{cozzo2013contact,wang2013effect}, and it is possible to observe localization phenomenons~\cite{de2017disease}. For susceptible-infected-recovered (SIR) model, Dickison, Havlin and Stanley~\cite{dickison2012epidemics} found that the system might exhibit a mixed phase, i.e., the epidemic outbreaks in one layer but not others. Refs.~\cite{wang2014asymmetrically,wang2016suppressing,
liu2016impacts} studied a model with the spreading of information and epidemics simultaneously. The studies found that the diffusion of information can inhibit the spreading of epidemics.

The inter-layer structure in multilayer networks has significant impacts on the dynamics. Parshani et al.~\cite{parshani2011inter} found that a positive inter-layer degree correlation will inhibit large scale cascading failures. Ref.~\cite{saumell2012epidemic} found that those degree correlations will make the epidemic outbreak more easily, while Ref.~\cite{wang2014asymmetrically} found that the positive inter-layer degree correlations can also inhibit the outbreak of the epidemic. Understanding what kind of interconnecting structure will lead to better performance for specific dynamics is an essential task for understanding dynamics on multilayer network and for designing better network structures. For spreading dynamics, when considering adding a single edge between layers, the optimal solution was given in Ref.~\cite{aguirre2013successful} analytically. Based on matrix perturbation theory, Ref.~\cite{aguirre2013successful} derive that connecting the two nodes with the largest eigenvector centralities in each layer minimizes the epidemic threshold while maximizes the spreading prevalence. For better synchronizability, Aguirre et al.~\cite{aguirre2014synchronization} studied the optimal strategy when adding one single edge between layers analytically. Based on matrix perturbation theory, they found that connecting large degree nodes will give better synchronizability. Li et al.~\cite{li2016synchronizability} further generalized the optimal strategy when adding two edges. When consider adding multiple interconnecting edges, current results are mostly based on numerical methods. Wei et al.~\cite{wei2018synchronizability} studied the interconnecting strategy numerically when adding a small number of edges for two-layer networks with random regular networks in each layer. Their studies suggest that adding inter-layer connections gives a more significant contribution to synchronizability compared to inner-layer connections. Wei et al.~\cite{wei2018maximizing} also did numerical simulations for the optimal strategy for general multilayer networks.

In this study, we investigate the problem of optimizing the spreading prevalence in two-layer networks by adding multiple inter-layer edges. For two isolated networks, we try to understand how to add a small number of edges to maximize the stationary spreading prevalence in the interconnected network. We mainly focus on the SIS model near the critical point. With a known formula given in~\cite{Goltsev2012}, the epidemic prevalence near the critical point can be written in terms of the leading eigenvalue and eigenvector.  We first develop a scheme for approximating the new leading eigenvalue and eigenvector for the interconnected network after adding those interconnecting edges. With this approximation scheme, we obtain a formula that predicts the stationary epidemic prevalence in the two-layer interconnected network. Then this approximated prevalence can be optimized among all possible inter-layer structures. The optimal inter-layer structure that maximizes the spreading prevalence will found to minimizes the spreading outbreak threshold for the interconnected network simultaneously. Numerical simulations are performed to compare the strategy with some other heuristic strategies. The proposed strategy gives a better performance at least near the critical point when adding a small number of interconnecting edges.

The rest of the paper is organized as follows. In Sec.~\ref{sec:model}, we introduce the basic setups of the model and some notations. In Sec.~\ref{sec:theory}, the theoretical derivations of the strategy is given. Then in Sec.~\ref{sec:results}, the strategy is tested and compared to some other heuristic strategies and finally in Sec.~\ref{sec:dis} we give some conclusions and discussions.

\section{Model} \label{sec:model}
Starting with two isolated networks $a$ and $b$, we add a fixed number of edges to interconnect the two networks. The way of adding the edges will affect the dynamical behaviors on the interconnected network. We focus on the way of adding these edges that maximizing the spreading prevalence.

Let the adjacency matrices of the networks $a,b$ be $G_a$ and $G_b$ respectively. The number of nodes in $a (b)$ is $N_a (N_b)$ and the number of edges is $M_a (M_b)$. The total number of nodes is denoted by $N=N_a+N_b$ and the total number of edges by $M=M_a+M_b$. The adjacency matrix of the combined network then is
\begin{equation}
G^0=\left(
{\begin{array}{cc}
G_a & 0 \\
0 & G_b
\end{array}}
\right).
\end{equation}

A set of edges with fixed cardinality $\delta M$ will be added between the two networks. After the operation the new network combining $a,b$ and interconnecting edges will have adjacency matrix $G=G^0+\delta G$, where
\begin{equation}
\delta G=\left(
{\begin{array}{cc}
0 & C \\
C^{\mathrm{T}} & 0
\end{array}}
\right).
\end{equation}
Here $C$ is an $N_a\times N_b$ matrix which indicates how the inter layer connections are added. Its elements take values $C_{ij}\in \{0,1\}$, where $C_{ij}=1$ if and edge is added between the $i$-th node of network $a$ and $j$-th node of network $b$ and $C_{ij}=0$ otherwise. The matrix satisfies the constraint on total number of added edges as
\begin{equation}
\left\langle\mathbf{1}_{N_a}, C\mathbf{1}_{N_b}\right\rangle=\delta M,
\end{equation}
where $\mathbf{1}_{N_a}=\left[1,\cdots,1\right]^T$ is all-one vector of length $N_a$ and $\langle\cdot \rangle$ is the inner product of two vectors.

By choosing among all possible assignment of $C$, the epidemic prevalence can be maximized. In this paper, we consider the SIS model. For the SIS model, each node is in either the susceptible or infected state. In a discrete-time setting, at time step $t$, infected nodes have a probability $\lambda$ to infect their susceptible neighbors independently. Then the infected nodes (not including nodes get infected at the current time step) become susceptible again with probability $\mu$. In large time limit, the density of infected nodes will converge to its stationary value. The target of the paper is to maximize the stationary spreading prevalence.

\section{Theory} \label{sec:theory}
We employ the discrete time Markov chain approach~\cite{gomez2010discrete} to describe SIS model on general networks. For the discrete time Markov chain approach, the status of a node $i$ is characterized by $\rho_i(t)$, which is the probability that $i$ is infected at time step $t$. Then $\rho_i(t)$ evolves according to the following discrete time equations
\begin{equation}\label{eq:mfEquation}
\rho_i(t+1)=(1-\mu)\rho_i(t)+\left(1-\rho_i(t)\right)\left(1-q_i(t)\right)
\end{equation}
for $i\in \{1,\cdots,N\}$ and
\begin{equation}\label{eq:q}
q_i(t)=\prod_{j=1}^N\left(1-\lambda G_{ij}\rho_j(t)\right).
\end{equation}
The first term on the r.h.s. of Eq.~(\ref{eq:mfEquation}) corresponds to the probability that node $i$ is infected at $t$ and not recovered in $t+1$, and second term is the probability that $i$ is susceptible at $t$ and get infected by at least one infected neighbor. $q_i(t)$ in Eq.~(\ref{eq:q}) is the probability that $i$ is not get infected by any of its neighbors at $t$.

The stationary solution is given by the limit in $t$
\begin{equation}
\rho_i=\lim_{t\to\infty}\rho_i(t).
\end{equation}
Let $\rho$ and $q$ be the vector with elements $\rho_i$ and $q_i$ for $i\in\{1,\cdots,N\}$. Near the critical point $\rho$ is expected to be small and the equation can be linearized as $q\approx 1-\lambda G \rho$, the stationary equation reads
\begin{equation}
G\rho=(\mu/\lambda)\rho.
\end{equation}
The spreading outbreaks only when $\lambda/\mu >1/\omega_1$ where $\omega_1$ is the leading eigenvalue of $G$.
It has been shown in~\cite{Goltsev2012} the stationary prevalence of epidemic is approximately
\begin{equation}\label{eq:rhoAprro}
\left\langle\rho\right\rangle\approx \left(\lambda^*\omega_1-1\right)\frac{\sum_{i=1}^N u_i }{N \sum_{i=1}^N u^3_i}
\end{equation}
where $\left\langle\rho\right\rangle=(1/N)\sum_{i=1}^N\rho_i$, $\lambda^*=\lambda/\mu$ is the effective infection probability and $u$ is the eigenvector corresponding to the eigenvalue $\omega_1$. It can be seen that $\left\langle\rho\right\rangle$ is determined by the leading eigenvalue and eigenvector approximately near the critical point. Since $G$ is obtained by adding a small number of edges to $G^0$, its spectra should be closely related to that of $G^0$. Next we develop a scheme to approximate the spectra of $G$ from that of $G^0$.

First consider the spectra of matrix $G^0$. Let the $\omega^a_k$ for $k\in \{1,\cdots,N_a\}$ be the $k$-th eigenvalue of $G_a$ and $v^a_k$ the corresponding eigenvector. Similarly $\omega^b_l$, $v^b_l$ with $l\in \{1,\cdots,N_b\}$ are eigenvalues and eigenvectors for $G_b$. The adjacency matrix of networks $a$ and $b$ combined $G^0$ is a diagonal block matrix by putting $G_a$ and $G_b$ in the diagonal, thus with eigenvalues
\begin{equation}
\{\omega^a_k:k=1,\cdots,N_a\}\cup \{\omega^b_l:l=1,\cdots,N_b\}.
\end{equation}
Clearly the leading eigenvalue of $G^0$ is $\max\{\omega^a_1,\omega^b_1\}$. Without losing of generality we assume $\omega^a_1\geq \omega^b_1$. The corresponding eigenvector for an eigenvalue $\omega^a_k$ is
\begin{equation}
\hat{v}^a_k=\left(v_k^a, 0 \right)^\mathrm{T},
\end{equation}
which is by combining $v^a_1$ and all-zero vector of length $N_b$. Similarly, for eigenvalue $\omega^b_l$, the corresponding eigenvector of $G^0$ is
\begin{equation}
\hat{v}^b_l=\left(0, v_l^b\right)^\mathrm{T}
\end{equation}
with zero vector of length $N_a$.

Now we consider adding a small number of interconnecting edges. By assuming these edges won't shift the spectra of the two isolated networks too much, a first approximation of $u$ would be proportional to $v_1^a$ in the first $N_a$ elements, and to $v_1^b$ in the last $N_b$ elements, thus written as
\begin{equation}
u^0=\beta^a\hat{v}^a_1+\beta^b\hat{v}^b_1
\end{equation}
where $\beta^a,\beta^b\in \mathbb{R}$ are coefficients to be determined. The eigenvector $u$ is given by the following limit
\begin{equation}
u= c \lim_{n\to \infty} \left(\omega_1^a\right)^{-n} G^n u^0
\end{equation}
for some constant $c$ and $G^n$ denotes self matrix multiplication of $G$ for $n$ times. Since the number of interconnecting edges is small, it can be assumed that $u^0$ is already close to $u$ by choosing $\beta^a$ and $\beta^b$ properly. The limit thus approximated by setting $n=1$ and this gives
\begin{equation}
u\approx\frac{c}{\omega^a_1} \left(\beta^a\omega^a_1\hat{v}^a_1+\beta^b\omega^b_1\hat{v}^b_1+\beta^a\delta G\hat{v}^a_1+\beta^b\delta G\hat{v}^b_1\right).
\end{equation}
Rescale the parameters by
\begin{equation}
c\beta^a \to \beta^a, \frac{c}{\omega^a_1}\beta^b\omega^b_1\to \beta^b,
\end{equation}
the eigenvector $u$ is approximated by the form
\begin{equation}
u\approx u^0+\delta u,
\end{equation}
where
\begin{equation}
\delta u=\frac{\beta^a}{\omega^a_1} \delta G \hat{v}^a_1+\frac{\beta^b}{\omega^b_1} \delta G \hat{v}^b_1.
\end{equation}
For scale-free networks which we mainly consider in the paper, it has been shown that the leading eigenvalue diverges in the thermodynamic limit~\cite{Goh2001}. Thus $\delta u$ can be ignored for large enough networks and $u$ is approximated as $u\approx u^0$.

Denote the gap by $g=\omega^a_1-\omega^b_1$. The leading eigenvalue $\omega_1$ of $G$ can be written as $\omega_1=\omega^a_1+\delta \omega_1$, which is $\omega^a_1$ plus a correction term $\delta \omega_1$. With these approximations we arrive at the following eigenvalue equation
\begin{equation}\label{eq:eigenApprox}
\left(G^0+\delta G\right)u^0=\left(\omega_1^a+\delta \omega_1 \right) u^0.
\end{equation}
By definition
\begin{equation}
G^0u^0=\beta^a\omega^a_1\hat{v}^a_1+\beta^b\omega^b_1\hat{v}^b_1,
\end{equation}
and after some algebra Eq.~(\ref{eq:eigenApprox}) becomes
\begin{equation}
\left(\delta G-\delta\omega_1 \mathbb{I}\right)u^0=\beta^b g \hat{v}^b_1,
\end{equation}
where $\mathbb{I}$ is the $N$ by $N$ identity matrix. Multiplying $\hat{v}^a_1$ and $\hat{v}^b_1$ from the left separately gives the following equations
\begin{equation}
\begin{split}
&\beta^b \left\langle v_1^a, C v_1^b \right\rangle=\delta \omega_1 \beta^a,\\
&\beta^a \left\langle v_1^a, C v_1^b \right\rangle-g \beta^b =\delta \omega_1 \beta^b.
\end{split}
\end{equation}
The equations can be written in the form of an eigenvalue equation
\begin{equation}
\left(
{\begin{array}{cc}
0 & E(C) \\
E(C) & -g
\end{array}}
\right)
\left(
{\begin{array}{c}
\beta^a \\
\beta^b
\end{array}}
\right)
=\delta \omega_1 \left(
{\begin{array}{c}
\beta^a \\
\beta^b
\end{array}}
\right)
\end{equation}
where we have denoted by $E(C)=\left\langle v_1^a, C v_1^b \right\rangle$. The equation gives two eigenvalues and pick the larger one which is
\begin{equation}\label{eq:domega}
\delta \omega_1=\frac{1}{2}\left(\sqrt{4E^2(C)+g^2}-g\right).
\end{equation}
The corresponding eigenvectors gives
\begin{equation}\label{eq:betas}
\begin{split}
&\beta^a=\sqrt{1+\left(\frac{g}{2E(C)}\right)^2}+\frac{g}{2E(C)},\\
&\beta^b=1.
\end{split}
\end{equation}
In summary, the leading eigenvalue and eigenvector of the interconnected network are approximated as
\begin{equation}\label{eq:appro}
\begin{split}
&\omega_1=\omega^a_1+\delta \omega_1,\\
&u=\beta^a \hat{v}^a_1+\beta^b \hat{v}_1^b,
\end{split}
\end{equation}
where $\delta\omega_1$, $\beta^a$ and $\beta^b$ are given in Eq.~(\ref{eq:domega}) and Eq.~(\ref{eq:betas}). Clearly when $g/E(C) \to 0$, $\beta^a\to 1$ and the two networks play equal role. Meanwhile if $g/E(C) \gg 0$, network $a$ dominates $b$.
Normalizing $u$ to unity and substituting back into~(\ref{eq:rhoAprro}) gives
\begin{equation}
\begin{split}
\left\langle\rho\right\rangle= &\left(\lambda^*\omega^a_1+\lambda^* \delta \omega_1-1\right) \\
&\times\frac{\left(\left(\beta^a\right)^3+\beta^a\right)\theta^a_1+\left(\left(\beta^a\right)^2+1\right) \theta^b_1}{N\left(\left(\beta^a\right)^3 \theta^a_3+\theta^b_3\right)}
\end{split}
\end{equation}
where
\begin{equation}
\theta^a_1=\left\langle \mathbf{1}_{N_a},v_1^a\right\rangle,\
\theta^a_3=\left\langle \mathbf{1}_{N_a},v_1^a\cdot v_1^a\cdot v_1^a\right\rangle
\end{equation}
and similarly for $\theta^b_1$, $\theta^b_3$.

With this approximation scheme, we can optimize $\left\langle\rho\right\rangle$ over $C$. Since $\left\langle\rho\right\rangle$ depends on the interconnecting matrix $C$ only in the form of $E(C)$, it is sufficient optimize $\left\langle\rho\right\rangle$ as a function of $E(C)$. After determining the optimal $E(C)$, we choose the matrix $C$ such that $E(C)$ is closest to its optimal value. Empirically, for all the networks considered in this study (see Sec.~\ref{sec:results}), $\left\langle\rho\right\rangle$ is an increasing function of $E(C)$ for small $E(C)$. Thus it is sufficient to choose $C$ such that $E(C)$ is maximized, or in other words, to choose the top $\delta M$ edges ranked by $v^a_1(i)\times v^b_1(j)$ for $i\in\{1,\cdots,N_a\},\ j\in \{1,\cdots,N_b\}$. Here $v^a_1(i)$ denotes the $i$th element of $v^a_1$ and similarly for $v^b_1(j)$. Note that a node can be connected to multiple nodes in the other layer. After determine $C$ in this way, we perform the check to ensure that $\left\langle\rho\right\rangle$ is non-decreasing in the region $(0,E(C)]$.  From Eq.~(\ref{eq:domega}), $\delta \omega_1$ is also an increasing function of $E(C)$. Therefore, the optimal strategy also maximizes $\delta \omega_1$ as well as $\left\langle\rho\right\rangle$. Since the outbreak threshold for the interconnected network is given by $1/\left(\omega^a_1+\delta \omega_1\right)$, as a consequence, the optimal inter-layer structure also minimizes the outbreak threshold. In the rest of the paper, the proposed strategy is called large eigenvector connecting (LEC), since it is by connecting inter-layer node pairs with larger product of the two nodes' eigenvector centrality.

\section{Results}\label{sec:results}
In this section we test the strategy proposed in Sec.~\ref{sec:theory} both for synthetic and real-world networks. For synthetic networks, we consider scale-free networks generated by uncorrelated configuration model (UCM). Denote by `SF $\alpha$' for scale-free network with degree distribution $p(k)\sim k^{-\alpha}$ and degree exponent $\alpha$, we consider the following three pairs of networks: (\romannumeral 1) SF $3.0$-SF $3.0$, (\romannumeral 2) SF $3.0$-SF $2.3$, (\romannumeral 3) SF $3.0$-SF $4.0$ and two pairs of real world networks: (\romannumeral 4) Advogato-Facebook, (\romannumeral 5) HepPh-HepTh.
Here Advogato~\cite{Massa2009} and Facebook~\cite{Leskovec2012} are two online social networks, and HepPh and HepTh~\cite{Leskovec2007} are citation networks of papers from the high energy physics-theory and high energy physics-phenomenology sections of the e-print arXiv. The networks are downloaded from~\cite{Konect}.  TABLE~\ref{table1} shows some basic statistics of the four real-world networks. When performing Monte-Carlo simulations on these networks, we first choose $\lceil 0.02N\rceil$ nodes as infected seeds. Then the spreading runs at discrete time steps. At each time step, all the infected nodes infect each of their susceptible neighbors with probability $\lambda$ independently and then returns to susceptible state with probability $\mu$. The steps are then repeated until the density of infected nodes in the network reaches the stationary value.

\begin{table}[htbp]
\caption{Some basic statistics of four real-world networks. The statistics includes the number of nodes ($N$), the number of edges ($M$), the maximal degree ($k_{\mathrm{max}}$), the first moment of the degree distribution ($\langle k \rangle$), the clustering coefficient ($c$), the theoretical epidemic threshold predicted by $\lambda_c^*=1/\omega_1$ and the modularity ($Q$)~\cite{newman2006modularity}.}
\begin{tabular}{cccccccc}
\toprule
\hline
\hline
\  & $N$ & $M$ & $k_{\mathrm{max}}$ & $\langle k \rangle$ & $c$ & $\lambda^*_c$ & $Q$\\
\midrule
\hline
Advogato & 5042 & 39227 &  803 & 15.56 & 0.092 & 0.014 & 0.337\\
Facebook & 2888 & 2981 & 769 &  2.0644 &   0.0004 & 0.036&  0.809\\
HepPh & 34401 &  420784 & 846 &  24.463 &  0.280 & 0.013 &  0.408\\
HepTh & 27400 &  352021 & 2468 & 25.695 & 0.269 & 0.009 &  0.328\\
\bottomrule
\hline
\hline
\end{tabular}
\label{table1}
\end{table}

\begin{figure*}
\begin{center}
\epsfig{file=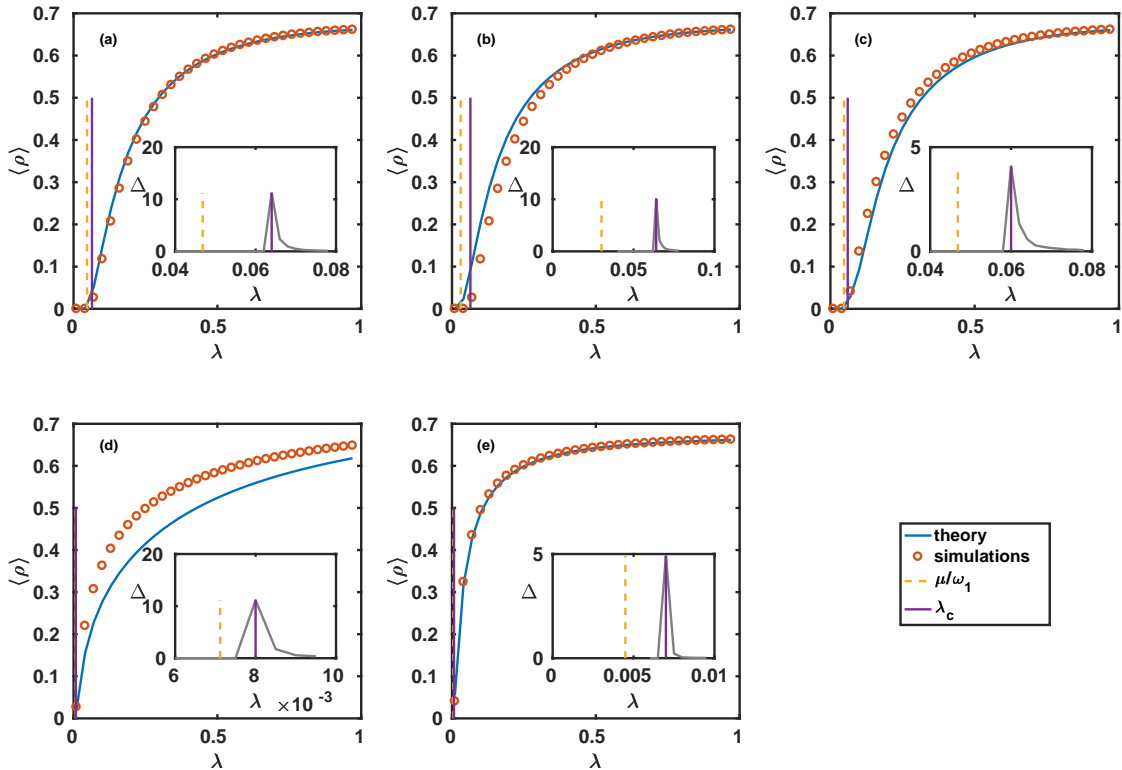,width=1\linewidth}
\caption{(Color online) The stationary epidemic size $\langle\rho\rangle$ as a function of the infection rate $\lambda$ with fixed recovery rate $\mu=0.5$ and inserted edge density $p=0.01$. $\lambda$ is changed from $0.01$ to $1$ at steps of $0.01$. The network pairs considered are (a) SF $3.0$-SF $3.0$, (b) SF $3.0$-SF $2.3$, (c) SF $3.0$-SF $4.0$, (d) Advogato-Facebook and (e) HepPh-HepTh. The orange circles correspond to the Monte Carlo simulations of $\langle\rho\rangle$ and the purple vertical lines denote the $\lambda_c$ estimated from simulations. The blue line is by iterating the Eqs.~(\ref{eq:mfEquation}), and the yellow dashed vertical line is the theoretical threshold. The gray line in the insets shows how the variability measure changes with $\lambda$ near the threshold. The variability measure is obtained by 500 independent runs of Monte Carlo simulations. Note that in the main plot of (d) and (e), $\lambda_c$ and $\mu/\omega_1$ might be too close to distinguish.}
\label{fig1}
\end{center}
\end{figure*}

First, we verify the accuracy of epidemic prevalence predicted by Eqs.~(\ref{eq:mfEquation}). Define the ratio $p=\delta M/M$. For the five network pairs considered, $\lceil pM \rceil$ of edges are added uniformly random between the two networks. Starting with random initial conditions, we iterate Eqs.~(\ref{eq:mfEquation}) until reaching a tolerated error. We perform Monte-Carlo simulations on the networks generated to compare with the theoretical predictions. Fig.~\ref{fig1} shows how $\langle\rho\rangle$ changes as a function of $\lambda$. The recovery rate is fixed to $\mu=0.5$ and as well in the rest of the paper. The density of interconnecting edges is $p=0.01$. In the insets of Fig.~\ref{fig1}, the theoretical epidemic threshold $\lambda_c^*=1/\omega_1$ is compared to the simulation results. The variability measure $\Delta$~\cite{Shu2015,Shu2016} is applied to determine the spreading threshold in simulations. In particular
\begin{equation}
\Delta=\sqrt{\overline{\langle \rho\rangle^2} /\overline{\langle \rho \rangle}^2-1},
\end{equation}
where the overline $\overline{\bullet}$ denotes the average over independent runs of the simulations. As a convention $\Delta=0$ when $\overline{\langle \rho \rangle}=0$. The simulated outbreak threshold is given by where the variability measure reaches its maximal. We mark the two thresholds by vertical lines both in Fig.~\ref{fig1} and the insets. From Fig.~\ref{fig1}, for all the network pairs considered expect Advogato-Facebook (Fig.~\ref{fig1}(d)), $\langle\rho\rangle$ and the epidemic threshold predicted by Eqs.~(\ref{eq:mfEquation}) agree with simulations in good accuracy. Meanwhile, for the network pair Advogato-Facebook, predictions of Eqs.~(\ref{eq:mfEquation}) are less accurate. Since mean-field theory ignores dynamical correlations~\cite{wang2017unification}, i.e., ignores dependencies among the states of each node and its neighbors, it has been shown that mean-field theory could be less accurate for real-world networks with high clustering coefficient~\cite{radicchi2016beyond}, high modularity~\cite{wang2016predicting} and low average degree~\cite{gleeson2012accuracy}. As shown in TABLE~\ref{table1}, the network Facebook has high modularity and low average degree, therefore, this could be the possible origins that Eqs.~(\ref{eq:mfEquation}) are less accurate for Advogato-Facebook. In general, quenched mean-field theory still predicts well the simulations, therefore can provide a necessary guarantee for the accuracy of the theory developed in Sec.~\ref{sec:theory}.

Next we test the accuracy of predictions for the leading eigenvalue $\omega_1$ and eigenvector $u$ of the interconnected network. We consecutively add interconnecting edges between the two networks according to the LEC strategy and compare the eigenvalue and eigenvector predicted by Eqs.~(\ref{eq:appro}) to the true values $\omega_1^*$ and $u^*$. Start with two isolated networks, we consecutively add the edges and do the comparison after each adding. To quantify the accuracy, for the leading eigenvalue we consider the relative error $\left(\omega_1^*-\omega_1\right)/\omega_1^*$, where $\omega_1^*$ is the true eigenvalue and $\omega_1$ is that predicted by Eqs.~(\ref{eq:appro}). For leading eigenvector the accuracy is measured by the cosine similarity between the true one $u^*$ and predicted one $u$. Let $\theta$ be the angle between $u^*$ and $u$,
\begin{equation}
\cos \theta = \left\langle u^*, u \right\rangle.
\end{equation}
Note that we have assumed both $u$ and $u^*$ are normalized to unity.
\begin{figure}
\begin{center}
\epsfig{file=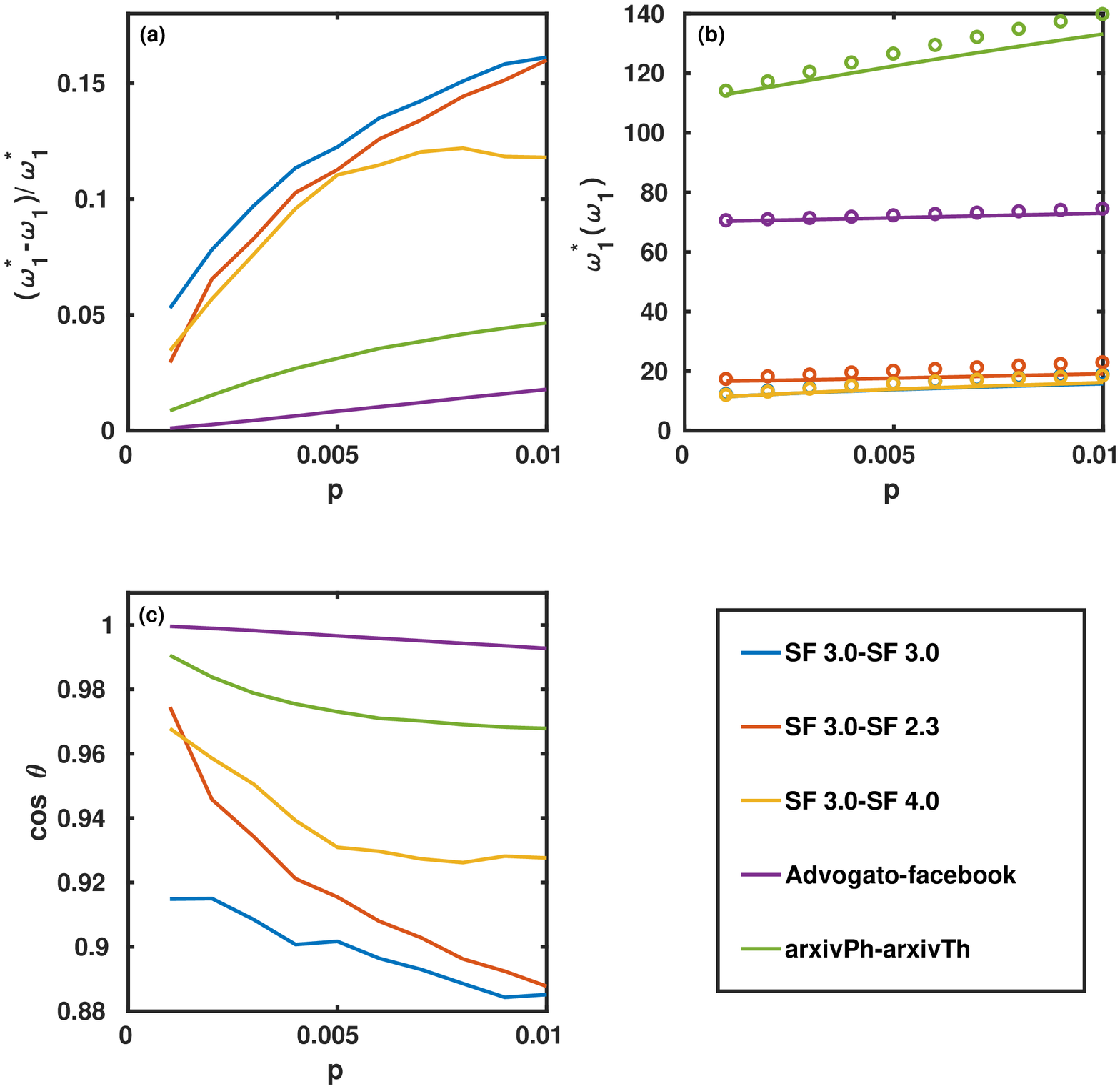,width=1\linewidth}
\caption{(Color online) The prediction accuracy of the leading eigenvalue and eigenvector of the interconnected network. The interconnected network is obtained by applying LEC to add edges between the two layers. (a) The relative error $\left(\omega_1^*-\omega_1\right)/\omega_1^*$ between the true eigenvalue $\omega_1^*$ and the theoretical prediction $\omega_1$ versus $p$. (b)  $\omega_1^*$ and $\omega_1$ versus $p$. (c) The cosine similarity between the true leading eigenvector and the theoretical prediction $u$. Different lines correspond to different network pairs. The fraction of edges added $p=\delta M/M$ starts from $0.001$ to $0.01$ at steps of $0.001$. In (b) the solid lines correspond to the theoretical predictions $\omega_1$ while the circles correspond to the true values $\omega_1^*$. In (b), the data points for SF $3.0$-SF $3.0$ and $3.0$-SF $4.0$ are highly overlapped.}
\label{fig2}
\end{center}
\end{figure}

Still denote the fraction of edges added by $p=\delta M/M$. $\left(\omega_1^*-\omega_1\right)/\omega_1^*$ and $\cos \theta$ versus $p$ are shown in Figs.~\ref{fig2}(a) and (c) respectively, while $\omega_1^*$ and $\omega_1$ versus $p$ are shown in Fig.~\ref{fig2}(b). From Figs.~\ref{fig2}, the theoretical predictions gives close approximations to it's true value. As can observed in Figs.~\ref{fig2}(a) and (b), errors in the eigenvalue $\omega_1^*-\omega_1$ have a small magnitude compared to the value of $\omega_1^*$, at least for small values of $p$. Also $\cos \theta$ is close to $1$ which indicates $u$ and $u^*$ are closely aligned. After tested the accuracy of Eqs.~(\ref{eq:mfEquation}) and Eqs.~(\ref{eq:appro}), the interconnecting strategy proposed in Sec.~\ref{sec:theory} is expected to be reliable.

Next we test the performance of the strategy of optimizing spreading. For comparison, the two following heuristic  strategies are considered. (\romannumeral 1) Large degree connecting (LDC). This is by choosing top candidate edges that ranked by the total degree of its two ends $k_i+k_j$. (\romannumeral 2) Random connecting (RC). The edges are chosen uniformly random among all node pairs. Remind that the strategy proposed in Sec.~\ref{sec:theory} is called largest eigenvector connecting (LEC).
\begin{figure*}
\begin{center}
\epsfig{file=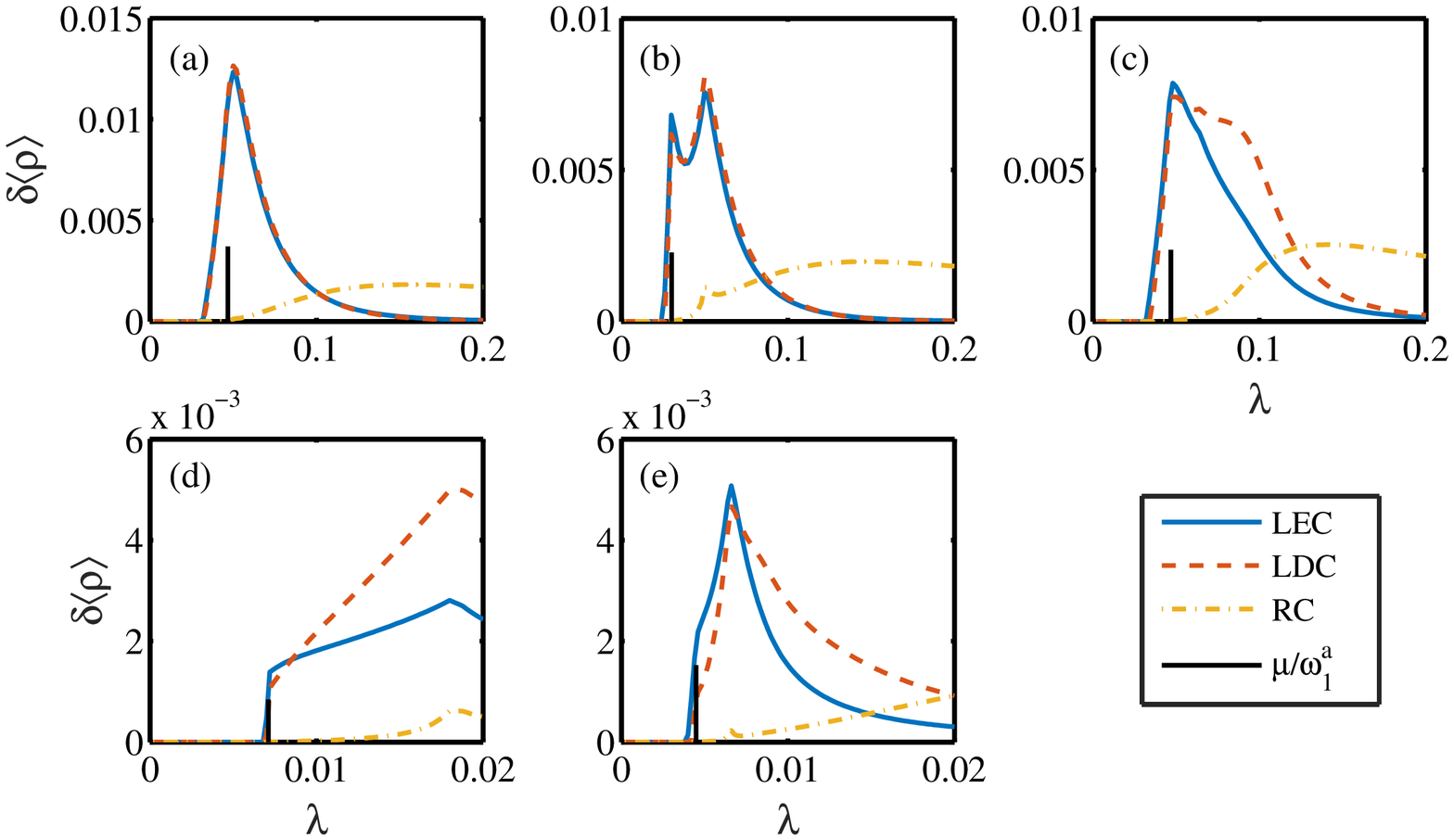,width=1\linewidth}
\caption{(Color online) $\delta \langle \rho \rangle$ versus $\lambda$ for $\mu=0.5$ and $p=0.005$ for the five pairs of networks (a) SF $3.0$-SF $3.0$, (b) SF $3.0$-SF $2.3$, (c) SF $3.0$-SF $4.0$, (d) Advogato-Facebook and (e) HepPh-HepTh. The values of $\lambda$ are chosen near the critical point thus depending on the specific network pairs considered. For the three synthetic network pairs, $\lambda$ is chosen from $0.002$ to $0.2$ with steps of $0.002$. For the real world networks $\lambda$ starts from $0.0002$ and ends at $0.02$ with steps of $0.0002$. The blue solid lines correspond to LEC, the orange dashed lines to LDC and the yellow dash-dot line to RC. The black vertical lines mark the critical value of $\lambda_c=\mu/\omega^a_1$ of the unconnected network $G^0$. }
\label{fig3}
\end{center}
\end{figure*}

As the model depends on the parameter $p$, we first consider comparing the three strategies with fixed $p=0.005$. The results for different small values of $p$ are similar as will be shown in the next. After choosing $C$ according to the strategies, Eqs.~(\ref{eq:mfEquation}) are iterated to stationary. For the convenience of visualization, the performance is evaluated by $\delta \langle \rho \rangle$, which is the difference of stationary $\langle \rho \rangle$ after and before adding the interconnecting edges in $C$. $\delta \langle \rho \rangle$ versus the infect rate $\lambda$ are shown in Fig.~\ref{fig3} for the five pairs of networks. Still, the recovery rate is fixed to $\mu=0.5$.  The threshold value of $\lambda_c=\mu/\omega^a_1$ for the isolated network $G^0$ is denoted by a black vertical line. For $\lambda$ near the critical point, LEC strategy gives highest $\delta \langle \rho \rangle$ for all the five pairs of networks. For the three model network pairs in Figs.~\ref{fig3}(a)-(c), LDC gives a very close performance to LEC. This is because that LEC and LDC predict similar $C$ for these three network pairs, as the degree centrality and eigenvalue centrality are strongly correlated in ranking for the UCM~\cite{lu2016vital}. For real-world networks, the structure becomes complex; therefore, degree and eigenvector centrality are less correlated. Thus a better performance of LEC is observed near the critical point for the two real-world network pairs.

As $\lambda$ becomes large and deviates the critical value, those nodes with large eigenvalue centrality have a high probability of to be infected. Thus interconnecting these nodes becomes unnecessary. Meanwhile, high degree nodes still play a central role in maintaining the epidemic. Thus gradually, better performance of the LDC strategy is observed. Eventually, when $\lambda$ become very large, the infected state is prevalent; therefore, all central nodes have a high probability to be infected. Meanwhile, the RC strategy more likely picks low degree nodes due to the power-law degree distribution. Thus in this region of $\lambda$, RC is expected to give the best performance among the three.

\begin{figure*}
\begin{center}
\epsfig{file=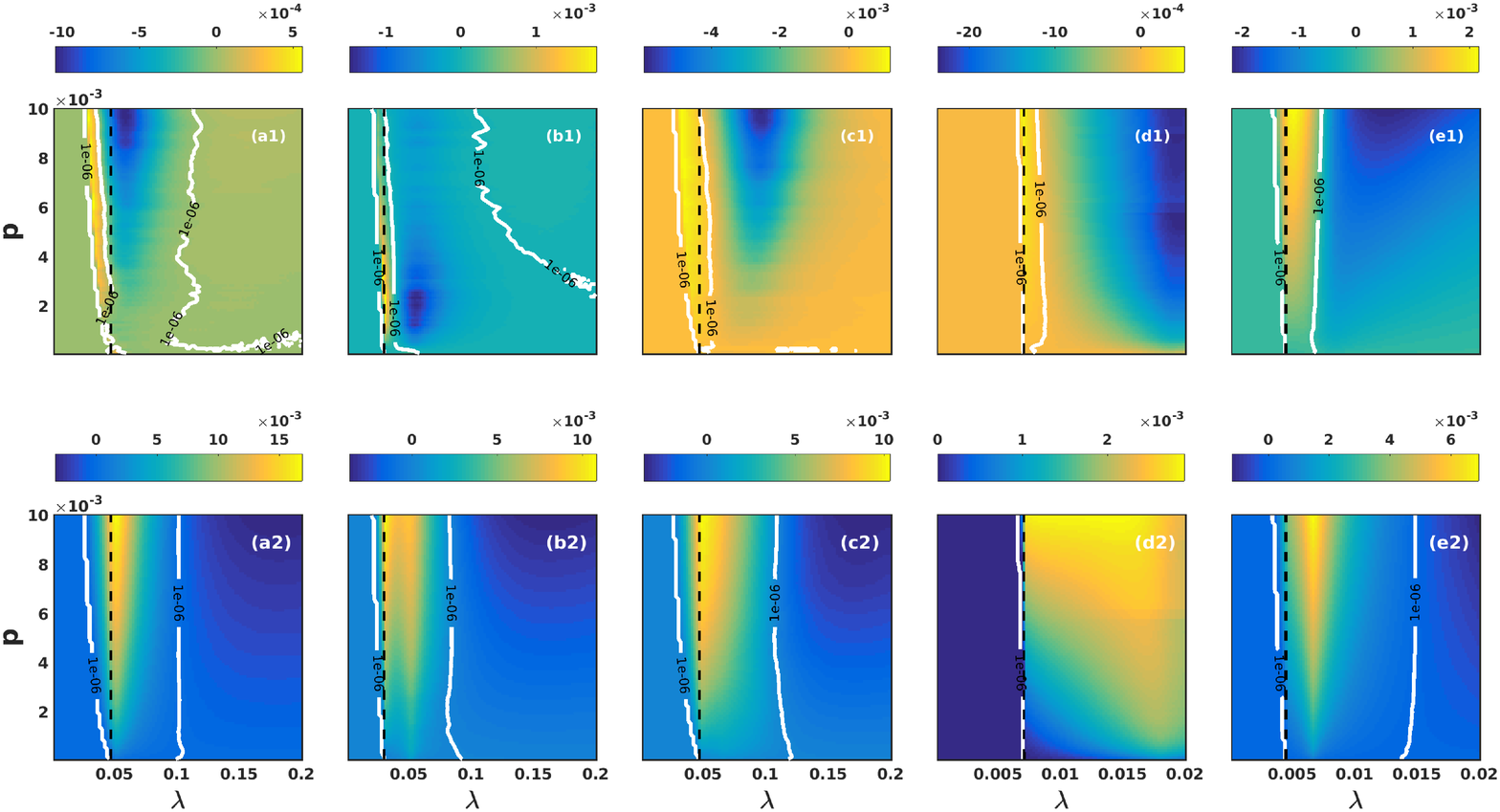,width=1\linewidth}
\caption{(Color online) The difference of $\langle \rho \rangle$ for LEC$-$LDC in the first row ((a1)-(e1)) and LEC$-$RC in second row ((a2)-(e2)). Each one of the five columns corresponds to one of the the network pairs, which are (a1-2) SF $3.0$-SF $3.0$, (b1-2) SF $3.0$-SF $2.3$, (c1-2) SF $3.0$-SF $4.0$, (d1-2) Advogato-Facebook and (e1-2) HepPh-HepTh. The fraction of added edges $p$ is changed from $0.0001$ to $0.01$ at steps of $0.0001$. The values of $\lambda$ are chosen close to the critical point and depends on specific network pairs. For the three synthetic network pairs, $\lambda$ is chosen from $0.002$ to $0.2$ with steps of $0.002$. For the real-world network pairs $\lambda$ starts from $0.0002$ and ends to $0.02$ at steps of $0.0002$. The while lines correspond to the contour of $\epsilon=1\mathrm{e}{-6}$, which bounds the region where LEC performs better than the other by at least $\epsilon$. The vertical black dashed lines indicate the position of $\lambda_c$ for the unconnected network.}
\label{fig4}
\end{center}
\end{figure*}

To get a more comprehensive picture, we study how the performance of the three strategies depend on $p$ and $\lambda$. For each point in the parameter space of $p$ and $\lambda$, we compute $\langle \rho \rangle$ for the three strategies and find the differences of LEC minus LDC and LEC minus RC. In other words, a positive difference indicates that LEC performs better than the other one compared. The results are shown in Fig.~\ref{fig4}. The first row in Fig.~\ref{fig4}, i.e. Fig.~\ref{fig4}(a1)-(e1), corresponds to LEC minus LDC for the five network pairs, and the second row (Figs.~\ref{fig4}(a2)-(e2)) correspond to LEC minus RC. The contours of $\epsilon=1\mathrm{e}{-6}$ are shown in the plots to indicate the region when LEC performs better. Here $\epsilon$ is not set to be $\epsilon=0$ to eliminate possible numeric errors. In Fig.~\ref{fig4}, the position of $\lambda_c=\mu/\omega^a_1$ for the unconnected network is marked by a vertical black dashed line. For all the cases in Fig.~\ref{fig4}, there is a vertical band in the parameter space where LEC gives better performance. The vertical band locates in where $\lambda$ is close to the critical value. Thus we can conclude that LEC is better than the other two near the critical point when the number of added edges is small.

\section{Discussion} \label{sec:dis}
In the study, we investigated the optimal strategy for interconnecting two isolated networks in order to maximize the spreading prevalence of SIS model. We develop a scheme to approximate the leading eigenvalue and eigenvector for the interconnected network. This approximation gives a formula that predicts the epidemic prevalence for the interconnected network. By maximizing the spreading prevalence over the interconnecting matrix, we obtain the optimal inter-layer structure. By numerically iterating the discrete-time Markov equations, we find the strategy gives better performance than some other heuristic strategies.

By maximizing the spreading prevalence $\langle \rho \rangle$ among all interconnecting matrices, it turns out that the strategy is equivalent to select edge candidates that are top-ranked by the product of eigenvector centrality of nodes in its two ends. Actually, for the networks considered in this study, $\langle \rho \rangle$ is not always an increasing function of $E(C)$ for large $E(C)$ and large $\lambda$. Nevertheless, in this region, the number of edges added must be very large, and the infection probability is far from critical; as a consequence, the eigenvalue approximation scheme becomes inaccurate. What is a better strategy when adding a large number of edges cannot be answered by the current approach.

Besides, the current method mainly focuses on the region near the epidemic threshold. In summary, we mainly work in regions where linear approximations are reliable. For larger effective spreading rates and larger numbers of added edges, the problem might become nonlinear and more complex. If it is possible to get a more comprehensive theory in all parameter regions requires further studies.

\begin{acknowledgments}
This work was partially supported by National Natural Science Foundation of China (Nos. 61673150, 11622538 and 61673086), China Postdoctoral Science Foundation (No. 2018M631073) and
the Fundamental Research Funds for the Central Universities.
\end{acknowledgments}


\begin{thebibliography}{59}%
\makeatletter
\providecommand \@ifxundefined [1]{%
 \@ifx{#1\undefined}
}%
\providecommand \@ifnum [1]{%
 \ifnum #1\expandafter \@firstoftwo
 \else \expandafter \@secondoftwo
 \fi
}%
\providecommand \@ifx [1]{%
 \ifx #1\expandafter \@firstoftwo
 \else \expandafter \@secondoftwo
 \fi
}%
\providecommand \natexlab [1]{#1}%
\providecommand \enquote  [1]{``#1''}%
\providecommand \bibnamefont  [1]{#1}%
\providecommand \bibfnamefont [1]{#1}%
\providecommand \citenamefont [1]{#1}%
\providecommand \href@noop [0]{\@secondoftwo}%
\providecommand \href [0]{\begingroup \@sanitize@url \@href}%
\providecommand \@href[1]{\@@startlink{#1}\@@href}%
\providecommand \@@href[1]{\endgroup#1\@@endlink}%
\providecommand \@sanitize@url [0]{\catcode `\\12\catcode `\$12\catcode
  `\&12\catcode `\#12\catcode `\^12\catcode `\_12\catcode `\%12\relax}%
\providecommand \@@startlink[1]{}%
\providecommand \@@endlink[0]{}%
\providecommand \url  [0]{\begingroup\@sanitize@url \@url }%
\providecommand \@url [1]{\endgroup\@href {#1}{\urlprefix }}%
\providecommand \urlprefix  [0]{URL }%
\providecommand \Eprint [0]{\href }%
\providecommand \doibase [0]{http://dx.doi.org/}%
\providecommand \selectlanguage [0]{\@gobble}%
\providecommand \bibinfo  [0]{\@secondoftwo}%
\providecommand \bibfield  [0]{\@secondoftwo}%
\providecommand \translation [1]{[#1]}%
\providecommand \BibitemOpen [0]{}%
\providecommand \bibitemStop [0]{}%
\providecommand \bibitemNoStop [0]{.\EOS\space}%
\providecommand \EOS [0]{\spacefactor3000\relax}%
\providecommand \BibitemShut  [1]{\csname bibitem#1\endcsname}%
\let\auto@bib@innerbib\@empty
\bibitem [{\citenamefont {Boccaletti}\ \emph {et~al.}(2014)\citenamefont
  {Boccaletti}, \citenamefont {Bianconi}, \citenamefont {Criado}, \citenamefont
  {Del~Genio}, \citenamefont {G{\'o}mez-Garde{\~n}es}, \citenamefont {Romance},
  \citenamefont {Sendi{\~n}a-Nadal}, \citenamefont {Wang},\ and\ \citenamefont
  {Zanin}}]{boccaletti2014structure}%
  \BibitemOpen
  \bibfield  {author} {\bibinfo {author} {\bibfnamefont {S.}~\bibnamefont
  {Boccaletti}}, \bibinfo {author} {\bibfnamefont {G.}~\bibnamefont
  {Bianconi}}, \bibinfo {author} {\bibfnamefont {R.}~\bibnamefont {Criado}},
  \bibinfo {author} {\bibfnamefont {C.~I.}\ \bibnamefont {Del~Genio}}, \bibinfo
  {author} {\bibfnamefont {J.}~\bibnamefont {G{\'o}mez-Garde{\~n}es}}, \bibinfo
  {author} {\bibfnamefont {M.}~\bibnamefont {Romance}}, \bibinfo {author}
  {\bibfnamefont {I.}~\bibnamefont {Sendi{\~n}a-Nadal}}, \bibinfo {author}
  {\bibfnamefont {Z.}~\bibnamefont {Wang}}, \ and\ \bibinfo {author}
  {\bibfnamefont {M.}~\bibnamefont {Zanin}},\ }\href@noop {} {\bibfield
  {journal} {\bibinfo  {journal} {Physics Reports}\ }\textbf {\bibinfo {volume}
  {544}},\ \bibinfo {pages} {1} (\bibinfo {year} {2014})}\BibitemShut {NoStop}%
\bibitem [{\citenamefont {Gao}\ \emph {et~al.}(2012)\citenamefont {Gao},
  \citenamefont {Buldyrev}, \citenamefont {Stanley},\ and\ \citenamefont
  {Havlin}}]{gao2012networks}%
  \BibitemOpen
  \bibfield  {author} {\bibinfo {author} {\bibfnamefont {J.}~\bibnamefont
  {Gao}}, \bibinfo {author} {\bibfnamefont {S.~V.}\ \bibnamefont {Buldyrev}},
  \bibinfo {author} {\bibfnamefont {H.~E.}\ \bibnamefont {Stanley}}, \ and\
  \bibinfo {author} {\bibfnamefont {S.}~\bibnamefont {Havlin}},\ }\href@noop {}
  {\bibfield  {journal} {\bibinfo  {journal} {Nature Physics}\ }\textbf
  {\bibinfo {volume} {8}},\ \bibinfo {pages} {40} (\bibinfo {year}
  {2012})}\BibitemShut {NoStop}%
\bibitem [{\citenamefont {Kivel{\"a}}\ \emph {et~al.}(2014)\citenamefont
  {Kivel{\"a}}, \citenamefont {Arenas}, \citenamefont {Barthelemy},
  \citenamefont {Gleeson}, \citenamefont {Moreno},\ and\ \citenamefont
  {Porter}}]{kivela2014multilayer}%
  \BibitemOpen
  \bibfield  {author} {\bibinfo {author} {\bibfnamefont {M.}~\bibnamefont
  {Kivel{\"a}}}, \bibinfo {author} {\bibfnamefont {A.}~\bibnamefont {Arenas}},
  \bibinfo {author} {\bibfnamefont {M.}~\bibnamefont {Barthelemy}}, \bibinfo
  {author} {\bibfnamefont {J.~P.}\ \bibnamefont {Gleeson}}, \bibinfo {author}
  {\bibfnamefont {Y.}~\bibnamefont {Moreno}}, \ and\ \bibinfo {author}
  {\bibfnamefont {M.~A.}\ \bibnamefont {Porter}},\ }\href@noop {} {\bibfield
  {journal} {\bibinfo  {journal} {Journal of Complex Networks}\ }\textbf
  {\bibinfo {volume} {2}},\ \bibinfo {pages} {203} (\bibinfo {year}
  {2014})}\BibitemShut {NoStop}%
\bibitem [{\citenamefont {De~Domenico}\ \emph {et~al.}(2016)\citenamefont
  {De~Domenico}, \citenamefont {Granell}, \citenamefont {Porter},\ and\
  \citenamefont {Arenas}}]{de2016physics}%
  \BibitemOpen
  \bibfield  {author} {\bibinfo {author} {\bibfnamefont {M.}~\bibnamefont
  {De~Domenico}}, \bibinfo {author} {\bibfnamefont {C.}~\bibnamefont
  {Granell}}, \bibinfo {author} {\bibfnamefont {M.~A.}\ \bibnamefont {Porter}},
  \ and\ \bibinfo {author} {\bibfnamefont {A.}~\bibnamefont {Arenas}},\
  }\href@noop {} {\bibfield  {journal} {\bibinfo  {journal} {Nature Physics}\
  }\textbf {\bibinfo {volume} {12}},\ \bibinfo {pages} {901} (\bibinfo {year}
  {2016})}\BibitemShut {NoStop}%
\bibitem [{\citenamefont {Cai}\ \emph {et~al.}(2018)\citenamefont {Cai},
  \citenamefont {Wang}, \citenamefont {Cui},\ and\ \citenamefont
  {Stanley}}]{cai2018multiplex}%
  \BibitemOpen
  \bibfield  {author} {\bibinfo {author} {\bibfnamefont {M.}~\bibnamefont
  {Cai}}, \bibinfo {author} {\bibfnamefont {W.}~\bibnamefont {Wang}}, \bibinfo
  {author} {\bibfnamefont {Y.}~\bibnamefont {Cui}}, \ and\ \bibinfo {author}
  {\bibfnamefont {H.~E.}\ \bibnamefont {Stanley}},\ }\href@noop {} {\bibfield
  {journal} {\bibinfo  {journal} {Physica A }\ }\textbf {\bibinfo {volume} {490}},\ \bibinfo {pages} {1}
  (\bibinfo {year} {2018})}\BibitemShut {NoStop}%
\bibitem [{\citenamefont {Sol{\'e}-Ribalta}\ \emph {et~al.}(2016)\citenamefont
  {Sol{\'e}-Ribalta}, \citenamefont {G{\'o}mez},\ and\ \citenamefont
  {Arenas}}]{sole2016congestion}%
  \BibitemOpen
  \bibfield  {author} {\bibinfo {author} {\bibfnamefont {A.}~\bibnamefont
  {Sol{\'e}-Ribalta}}, \bibinfo {author} {\bibfnamefont {S.}~\bibnamefont
  {G{\'o}mez}}, \ and\ \bibinfo {author} {\bibfnamefont {A.}~\bibnamefont
  {Arenas}},\ }\href@noop {} {\bibfield  {journal} {\bibinfo  {journal}
  {Physical Review Letters}\ }\textbf {\bibinfo {volume} {116}},\ \bibinfo
  {pages} {108701} (\bibinfo {year} {2016})}\BibitemShut {NoStop}%
\bibitem [{\citenamefont {Buono}\ and\ \citenamefont
  {Braunstein}(2015)}]{buono2015immunization}%
  \BibitemOpen
  \bibfield  {author} {\bibinfo {author} {\bibfnamefont {C.}~\bibnamefont
  {Buono}}\ and\ \bibinfo {author} {\bibfnamefont {L.~A.}\ \bibnamefont
  {Braunstein}},\ }\href@noop {} {\bibfield  {journal} {\bibinfo  {journal}
  {Europhysics Letters}\ }\textbf {\bibinfo {volume} {109}},\ \bibinfo {pages}
  {26001} (\bibinfo {year} {2015})}\BibitemShut {NoStop}%
\bibitem [{\citenamefont {Buldyrev}\ \emph {et~al.}(2010)\citenamefont
  {Buldyrev}, \citenamefont {Parshani}, \citenamefont {Paul}, \citenamefont
  {Stanley},\ and\ \citenamefont {Havlin}}]{buldyrev2010catastrophic}%
  \BibitemOpen
  \bibfield  {author} {\bibinfo {author} {\bibfnamefont {S.~V.}\ \bibnamefont
  {Buldyrev}}, \bibinfo {author} {\bibfnamefont {R.}~\bibnamefont {Parshani}},
  \bibinfo {author} {\bibfnamefont {G.}~\bibnamefont {Paul}}, \bibinfo {author}
  {\bibfnamefont {H.~E.}\ \bibnamefont {Stanley}}, \ and\ \bibinfo {author}
  {\bibfnamefont {S.}~\bibnamefont {Havlin}},\ }\href@noop {} {\bibfield
  {journal} {\bibinfo  {journal} {Nature}\ }\textbf {\bibinfo {volume} {464}},\
  \bibinfo {pages} {1025} (\bibinfo {year} {2010})}\BibitemShut {NoStop}%
\bibitem [{\citenamefont {Cozzo}\ \emph {et~al.}(2013)\citenamefont {Cozzo},
  \citenamefont {Banos}, \citenamefont {Meloni},\ and\ \citenamefont
  {Moreno}}]{cozzo2013contact}%
  \BibitemOpen
  \bibfield  {author} {\bibinfo {author} {\bibfnamefont {E.}~\bibnamefont
  {Cozzo}}, \bibinfo {author} {\bibfnamefont {R.~A.}\ \bibnamefont {Banos}},
  \bibinfo {author} {\bibfnamefont {S.}~\bibnamefont {Meloni}}, \ and\ \bibinfo
  {author} {\bibfnamefont {Y.}~\bibnamefont {Moreno}},\ }\href@noop {}
  {\bibfield  {journal} {\bibinfo  {journal} {Physical Review E}\ }\textbf
  {\bibinfo {volume} {88}},\ \bibinfo {pages} {050801} (\bibinfo {year}
  {2013})}\BibitemShut {NoStop}%
\bibitem [{\citenamefont {Kim}\ and\ \citenamefont
  {Goh}(2013)}]{kim2013coevolution}%
  \BibitemOpen
  \bibfield  {author} {\bibinfo {author} {\bibfnamefont {J.~Y.}\ \bibnamefont
  {Kim}}\ and\ \bibinfo {author} {\bibfnamefont {K.-I.}\ \bibnamefont {Goh}},\
  }\href@noop {} {\bibfield  {journal} {\bibinfo  {journal} {Physical Review
  Letters}\ }\textbf {\bibinfo {volume} {111}},\ \bibinfo {pages} {058702}
  (\bibinfo {year} {2013})}\BibitemShut {NoStop}%
\bibitem [{\citenamefont {Lee}\ \emph {et~al.}(2012)\citenamefont {Lee},
  \citenamefont {Kim}, \citenamefont {Cho}, \citenamefont {Goh},\ and\
  \citenamefont {Kim}}]{lee2012correlated}%
  \BibitemOpen
  \bibfield  {author} {\bibinfo {author} {\bibfnamefont {K.-M.}\ \bibnamefont
  {Lee}}, \bibinfo {author} {\bibfnamefont {J.~Y.}\ \bibnamefont {Kim}},
  \bibinfo {author} {\bibfnamefont {W.-k.}\ \bibnamefont {Cho}}, \bibinfo
  {author} {\bibfnamefont {K.-I.}\ \bibnamefont {Goh}}, \ and\ \bibinfo
  {author} {\bibfnamefont {I.}~\bibnamefont {Kim}},\ }\href@noop {} {\bibfield
  {journal} {\bibinfo  {journal} {New Journal of Physics}\ }\textbf {\bibinfo
  {volume} {14}},\ \bibinfo {pages} {033027} (\bibinfo {year}
  {2012})}\BibitemShut {NoStop}%
\bibitem [{\citenamefont {Wang}\ \emph {et~al.}(2015)\citenamefont {Wang},
  \citenamefont {Wang}, \citenamefont {Szolnoki},\ and\ \citenamefont
  {Perc}}]{wang2015evolutionary}%
  \BibitemOpen
  \bibfield  {author} {\bibinfo {author} {\bibfnamefont {Z.}~\bibnamefont
  {Wang}}, \bibinfo {author} {\bibfnamefont {L.}~\bibnamefont {Wang}}, \bibinfo
  {author} {\bibfnamefont {A.}~\bibnamefont {Szolnoki}}, \ and\ \bibinfo
  {author} {\bibfnamefont {M.}~\bibnamefont {Perc}},\ }\href@noop {} {\bibfield
   {journal} {\bibinfo  {journal} {European Physical Journal B}\ }\textbf
  {\bibinfo {volume} {88}} (\bibinfo {year} {2015})},\ \Eprint
  {http://arxiv.org/abs/1504.04359} {1504.04359} \BibitemShut {NoStop}%
\bibitem [{\citenamefont {Zhang}\ \emph {et~al.}(2015)\citenamefont {Zhang},
  \citenamefont {Boccaletti}, \citenamefont {Guan},\ and\ \citenamefont
  {Liu}}]{zhang2015explosive}%
  \BibitemOpen
  \bibfield  {author} {\bibinfo {author} {\bibfnamefont {X.}~\bibnamefont
  {Zhang}}, \bibinfo {author} {\bibfnamefont {S.}~\bibnamefont {Boccaletti}},
  \bibinfo {author} {\bibfnamefont {S.}~\bibnamefont {Guan}}, \ and\ \bibinfo
  {author} {\bibfnamefont {Z.}~\bibnamefont {Liu}},\ }\href@noop {} {\bibfield
  {journal} {\bibinfo  {journal} {Physical Review Letters}\ }\textbf {\bibinfo
  {volume} {114}},\ \bibinfo {pages} {038701} (\bibinfo {year}
  {2015})}\BibitemShut {NoStop}%
\bibitem [{\citenamefont {Valdez}\ \emph {et~al.}(2013)\citenamefont {Valdez},
  \citenamefont {Macri}, \citenamefont {Stanley},\ and\ \citenamefont
  {Braunstein}}]{valdez2013triple}%
  \BibitemOpen
  \bibfield  {author} {\bibinfo {author} {\bibfnamefont {L.~D.}\ \bibnamefont
  {Valdez}}, \bibinfo {author} {\bibfnamefont {P.~A.}\ \bibnamefont {Macri}},
  \bibinfo {author} {\bibfnamefont {H.~E.}\ \bibnamefont {Stanley}}, \ and\
  \bibinfo {author} {\bibfnamefont {L.~A.}\ \bibnamefont {Braunstein}},\
  }\href@noop {} {\bibfield  {journal} {\bibinfo  {journal} {Physical Review
  E}\ }\textbf {\bibinfo {volume} {88}},\ \bibinfo {pages} {050803} (\bibinfo
  {year} {2013})}\BibitemShut {NoStop}%
\bibitem [{\citenamefont {Di~Muro}\ \emph
  {et~al.}(2016{\natexlab{a}})\citenamefont {Di~Muro}, \citenamefont
  {La~Rocca}, \citenamefont {Stanley}, \citenamefont {Havlin},\ and\
  \citenamefont {Braunstein}}]{di2016recovery}%
  \BibitemOpen
  \bibfield  {author} {\bibinfo {author} {\bibfnamefont {M.}~\bibnamefont
  {Di~Muro}}, \bibinfo {author} {\bibfnamefont {C.}~\bibnamefont {La~Rocca}},
  \bibinfo {author} {\bibfnamefont {H.}~\bibnamefont {Stanley}}, \bibinfo
  {author} {\bibfnamefont {S.}~\bibnamefont {Havlin}}, \ and\ \bibinfo {author}
  {\bibfnamefont {L.}~\bibnamefont {Braunstein}},\ }\href@noop {} {\bibfield
  {journal} {\bibinfo  {journal} {Scientific Reports}\ }\textbf {\bibinfo
  {volume} {6}},\ \bibinfo {pages} {22834} (\bibinfo {year}
  {2016}{\natexlab{a}})}\BibitemShut {NoStop}%
\bibitem [{\citenamefont {Di~Muro}\ \emph
  {et~al.}(2016{\natexlab{b}})\citenamefont {Di~Muro}, \citenamefont
  {Buldyrev}, \citenamefont {Stanley},\ and\ \citenamefont
  {Braunstein}}]{di2016cascading}%
  \BibitemOpen
  \bibfield  {author} {\bibinfo {author} {\bibfnamefont {M.}~\bibnamefont
  {Di~Muro}}, \bibinfo {author} {\bibfnamefont {S.}~\bibnamefont {Buldyrev}},
  \bibinfo {author} {\bibfnamefont {H.}~\bibnamefont {Stanley}}, \ and\
  \bibinfo {author} {\bibfnamefont {L.}~\bibnamefont {Braunstein}},\
  }\href@noop {} {\bibfield  {journal} {\bibinfo  {journal} {Physical Review
  E}\ }\textbf {\bibinfo {volume} {94}},\ \bibinfo {pages} {042304} (\bibinfo
  {year} {2016}{\natexlab{b}})}\BibitemShut {NoStop}%
\bibitem [{\citenamefont {Valdez}\ \emph {et~al.}(2016)\citenamefont {Valdez},
  \citenamefont {Di~Muro},\ and\ \citenamefont
  {Braunstein}}]{valdez2016failure}%
  \BibitemOpen
  \bibfield  {author} {\bibinfo {author} {\bibfnamefont {L.}~\bibnamefont
  {Valdez}}, \bibinfo {author} {\bibfnamefont {M.}~\bibnamefont {Di~Muro}}, \
  and\ \bibinfo {author} {\bibfnamefont {L.}~\bibnamefont {Braunstein}},\
  }\href@noop {} {\bibfield  {journal} {\bibinfo  {journal} {Journal of
  Statistical Mechanics: Theory and Experiment}\ }\textbf {\bibinfo {volume}
  {2016}},\ \bibinfo {pages} {093402} (\bibinfo {year} {2016})}\BibitemShut
  {NoStop}%
\bibitem [{\citenamefont {Cohen}\ \emph {et~al.}(2002)\citenamefont {Cohen},
  \citenamefont {Ben-Avraham},\ and\ \citenamefont
  {Havlin}}]{cohen2002percolation}%
  \BibitemOpen
  \bibfield  {author} {\bibinfo {author} {\bibfnamefont {R.}~\bibnamefont
  {Cohen}}, \bibinfo {author} {\bibfnamefont {D.}~\bibnamefont {Ben-Avraham}},
  \ and\ \bibinfo {author} {\bibfnamefont {S.}~\bibnamefont {Havlin}},\
  }\href@noop {} {\bibfield  {journal} {\bibinfo  {journal} {Physical Review
  E}\ }\textbf {\bibinfo {volume} {66}},\ \bibinfo {pages} {036113} (\bibinfo
  {year} {2002})}\BibitemShut {NoStop}%
\bibitem [{\citenamefont {Serrano}\ and\ \citenamefont
  {Bogun{\'a}}(2006)}]{serrano2006percolation}%
  \BibitemOpen
  \bibfield  {author} {\bibinfo {author} {\bibfnamefont {M.~{\'A}.}\
  \bibnamefont {Serrano}}\ and\ \bibinfo {author} {\bibfnamefont
  {M.}~\bibnamefont {Bogun{\'a}}},\ }\href@noop {} {\bibfield  {journal}
  {\bibinfo  {journal} {Physical Review Letters}\ }\textbf {\bibinfo {volume}
  {97}},\ \bibinfo {pages} {088701} (\bibinfo {year} {2006})}\BibitemShut
  {NoStop}%
\bibitem [{\citenamefont {Goltsev}\ \emph {et~al.}(2008)\citenamefont
  {Goltsev}, \citenamefont {Dorogovtsev},\ and\ \citenamefont
  {Mendes}}]{goltsev2008percolation}%
  \BibitemOpen
  \bibfield  {author} {\bibinfo {author} {\bibfnamefont {A.}~\bibnamefont
  {Goltsev}}, \bibinfo {author} {\bibfnamefont {S.}~\bibnamefont
  {Dorogovtsev}}, \ and\ \bibinfo {author} {\bibfnamefont {J.}~\bibnamefont
  {Mendes}},\ }\href@noop {} {\bibfield  {journal} {\bibinfo  {journal}
  {Physical Review E}\ }\textbf {\bibinfo {volume} {78}},\ \bibinfo {pages}
  {051105} (\bibinfo {year} {2008})}\BibitemShut {NoStop}%
\bibitem [{\citenamefont {Perc}\ \emph {et~al.}(2017)\citenamefont {Perc},
  \citenamefont {Jordan}, \citenamefont {Rand}, \citenamefont {Wang},
  \citenamefont {Boccaletti},\ and\ \citenamefont
  {Szolnoki}}]{perc2017statistical}%
  \BibitemOpen
  \bibfield  {author} {\bibinfo {author} {\bibfnamefont {M.}~\bibnamefont
  {Perc}}, \bibinfo {author} {\bibfnamefont {J.~J.}\ \bibnamefont {Jordan}},
  \bibinfo {author} {\bibfnamefont {D.~G.}\ \bibnamefont {Rand}}, \bibinfo
  {author} {\bibfnamefont {Z.}~\bibnamefont {Wang}}, \bibinfo {author}
  {\bibfnamefont {S.}~\bibnamefont {Boccaletti}}, \ and\ \bibinfo {author}
  {\bibfnamefont {A.}~\bibnamefont {Szolnoki}},\ }\href@noop {} {\bibfield
  {journal} {\bibinfo  {journal} {Physics Reports}\ }\textbf {\bibinfo {volume}
  {687}},\ \bibinfo {pages} {1} (\bibinfo {year} {2017})}\BibitemShut {NoStop}%
\bibitem [{\citenamefont {Wang}\ \emph
  {et~al.}(2013{\natexlab{a}})\citenamefont {Wang}, \citenamefont {Szolnoki},\
  and\ \citenamefont {Perc}}]{wang2013interdependent}%
  \BibitemOpen
  \bibfield  {author} {\bibinfo {author} {\bibfnamefont {Z.}~\bibnamefont
  {Wang}}, \bibinfo {author} {\bibfnamefont {A.}~\bibnamefont {Szolnoki}}, \
  and\ \bibinfo {author} {\bibfnamefont {M.}~\bibnamefont {Perc}},\ }\href@noop
  {} {\bibfield  {journal} {\bibinfo  {journal} {Scientific Reports}\ }\textbf
  {\bibinfo {volume} {3}} (\bibinfo {year} {2013}{\natexlab{a}})}\BibitemShut
  {NoStop}%
\bibitem [{\citenamefont {Wang}\ \emph {et~al.}(2019)\citenamefont {Wang},
  \citenamefont {Liu}, \citenamefont {Liang}, \citenamefont {Hu},\ and\
  \citenamefont {Zhou}}]{wang2019coevolution}%
  \BibitemOpen
  \bibfield  {author} {\bibinfo {author} {\bibfnamefont {W.}~\bibnamefont
  {Wang}}, \bibinfo {author} {\bibfnamefont {Q.-H.}\ \bibnamefont {Liu}},
  \bibinfo {author} {\bibfnamefont {J.}~\bibnamefont {Liang}}, \bibinfo
  {author} {\bibfnamefont {Y.}~\bibnamefont {Hu}}, \ and\ \bibinfo {author}
  {\bibfnamefont {T.}~\bibnamefont {Zhou}},\ }\href@noop {} {\bibfield
  {journal} {\bibinfo  {journal} {arXiv:1901.02125}\ } (\bibinfo {year}
  {2019})}\BibitemShut {NoStop}%
\bibitem [{\citenamefont {Brummitt}\ \emph {et~al.}(2012)\citenamefont
  {Brummitt}, \citenamefont {Lee},\ and\ \citenamefont
  {Goh}}]{brummitt2012multiplexity}%
  \BibitemOpen
  \bibfield  {author} {\bibinfo {author} {\bibfnamefont {C.~D.}\ \bibnamefont
  {Brummitt}}, \bibinfo {author} {\bibfnamefont {K.-M.}\ \bibnamefont {Lee}}, \
  and\ \bibinfo {author} {\bibfnamefont {K.-I.}\ \bibnamefont {Goh}},\
  }\href@noop {} {\bibfield  {journal} {\bibinfo  {journal} {Physical Review
  E}\ }\textbf {\bibinfo {volume} {85}},\ \bibinfo {pages} {045102} (\bibinfo
  {year} {2012})}\BibitemShut {NoStop}%
\bibitem [{\citenamefont {Lee}\ \emph {et~al.}(2014)\citenamefont {Lee},
  \citenamefont {Brummitt},\ and\ \citenamefont {Goh}}]{lee2014threshold}%
  \BibitemOpen
  \bibfield  {author} {\bibinfo {author} {\bibfnamefont {K.-M.}\ \bibnamefont
  {Lee}}, \bibinfo {author} {\bibfnamefont {C.~D.}\ \bibnamefont {Brummitt}}, \
  and\ \bibinfo {author} {\bibfnamefont {K.-I.}\ \bibnamefont {Goh}},\
  }\href@noop {} {\bibfield  {journal} {\bibinfo  {journal} {Physical Review
  E}\ }\textbf {\bibinfo {volume} {90}},\ \bibinfo {pages} {062816} (\bibinfo
  {year} {2014})}\BibitemShut {NoStop}%
\bibitem [{\citenamefont {Ya\u{g}an}\ and\ \citenamefont
  {Gligor}(2012)}]{Yagan2012}%
  \BibitemOpen
  \bibfield  {author} {\bibinfo {author} {\bibfnamefont {O.}~\bibnamefont
  {Ya\u{g}an}}\ and\ \bibinfo {author} {\bibfnamefont {V.}~\bibnamefont
  {Gligor}},\ }\href@noop {} {\bibfield  {journal} {\bibinfo  {journal} {Phys.
  Rev. E}\ }\textbf {\bibinfo {volume} {86}},\ \bibinfo {pages} {036103}
  (\bibinfo {year} {2012})}\BibitemShut {NoStop}%
\bibitem [{\citenamefont {Wang}\ \emph
  {et~al.}(2018{\natexlab{a}})\citenamefont {Wang}, \citenamefont {Chen},\ and\
  \citenamefont {Zhong}}]{wang2018social}%
  \BibitemOpen
  \bibfield  {author} {\bibinfo {author} {\bibfnamefont {W.}~\bibnamefont
  {Wang}}, \bibinfo {author} {\bibfnamefont {X.-L.}\ \bibnamefont {Chen}}, \
  and\ \bibinfo {author} {\bibfnamefont {L.-F.}\ \bibnamefont {Zhong}},\
  }\href@noop {} {\bibfield  {journal} {\bibinfo  {journal} {Physica A }\ }\textbf {\bibinfo {volume}
  {503}},\ \bibinfo {pages} {604} (\bibinfo {year}
  {2018}{\natexlab{a}})}\BibitemShut {NoStop}%
\bibitem [{\citenamefont {Wang}\ \emph
  {et~al.}(2018{\natexlab{b}})\citenamefont {Wang}, \citenamefont {Tang},
  \citenamefont {Stanley},\ and\ \citenamefont {Braunstein}}]{wang2018social2}%
  \BibitemOpen
  \bibfield  {author} {\bibinfo {author} {\bibfnamefont {W.}~\bibnamefont
  {Wang}}, \bibinfo {author} {\bibfnamefont {M.}~\bibnamefont {Tang}}, \bibinfo
  {author} {\bibfnamefont {H.~E.}\ \bibnamefont {Stanley}}, \ and\ \bibinfo
  {author} {\bibfnamefont {L.~A.}\ \bibnamefont {Braunstein}},\ }\href@noop {}
  {\bibfield  {journal} {\bibinfo  {journal} {Physical Review E}\ }\textbf
  {\bibinfo {volume} {98}},\ \bibinfo {pages} {062320} (\bibinfo {year}
  {2018}{\natexlab{b}})}\BibitemShut {NoStop}%
\bibitem [{\citenamefont {Shu}\ \emph {et~al.}(2018)\citenamefont {Shu},
  \citenamefont {Liu}, \citenamefont {Wang},\ and\ \citenamefont
  {Wang}}]{shu2018social}%
  \BibitemOpen
  \bibfield  {author} {\bibinfo {author} {\bibfnamefont {P.}~\bibnamefont
  {Shu}}, \bibinfo {author} {\bibfnamefont {Q.-H.}\ \bibnamefont {Liu}},
  \bibinfo {author} {\bibfnamefont {S.}~\bibnamefont {Wang}}, \ and\ \bibinfo
  {author} {\bibfnamefont {W.}~\bibnamefont {Wang}},\ }\href@noop {} {\bibfield
   {journal} {\bibinfo  {journal} {Chaos }\ }\textbf {\bibinfo {volume} {28}},\ \bibinfo {pages}
  {113114} (\bibinfo {year} {2018})}\BibitemShut {NoStop}%
\bibitem [{\citenamefont {Chen}\ \emph {et~al.}(2018)\citenamefont {Chen},
  \citenamefont {Wang}, \citenamefont {Cai}, \citenamefont {Stanley},\ and\
  \citenamefont {Braunstein}}]{chen2018optimal}%
  \BibitemOpen
  \bibfield  {author} {\bibinfo {author} {\bibfnamefont {X.}~\bibnamefont
  {Chen}}, \bibinfo {author} {\bibfnamefont {W.}~\bibnamefont {Wang}}, \bibinfo
  {author} {\bibfnamefont {S.}~\bibnamefont {Cai}}, \bibinfo {author}
  {\bibfnamefont {H.~E.}\ \bibnamefont {Stanley}}, \ and\ \bibinfo {author}
  {\bibfnamefont {L.~A.}\ \bibnamefont {Braunstein}},\ }\href@noop {}
  {\bibfield  {journal} {\bibinfo  {journal} {Journal of Statistical Mechanics:
  Theory and Experiment}\ }\textbf {\bibinfo {volume} {2018}},\ \bibinfo
  {pages} {053501} (\bibinfo {year} {2018})}\BibitemShut {NoStop}%
\bibitem [{\citenamefont {Saumell-Mendiola}\ \emph {et~al.}(2012)\citenamefont
  {Saumell-Mendiola}, \citenamefont {Serrano},\ and\ \citenamefont
  {Bogun{\'a}}}]{saumell2012epidemic}%
  \BibitemOpen
  \bibfield  {author} {\bibinfo {author} {\bibfnamefont {A.}~\bibnamefont
  {Saumell-Mendiola}}, \bibinfo {author} {\bibfnamefont {M.~{\'A}.}\
  \bibnamefont {Serrano}}, \ and\ \bibinfo {author} {\bibfnamefont
  {M.}~\bibnamefont {Bogun{\'a}}},\ }\href@noop {} {\bibfield  {journal}
  {\bibinfo  {journal} {Physical Review E}\ }\textbf {\bibinfo {volume} {86}},\
  \bibinfo {pages} {026106} (\bibinfo {year} {2012})}\BibitemShut {NoStop}%
\bibitem [{\citenamefont {Pastor-Satorras}\ and\ \citenamefont
  {Vespignani}(2001)}]{pastor2001epidemic}%
  \BibitemOpen
  \bibfield  {author} {\bibinfo {author} {\bibfnamefont {R.}~\bibnamefont
  {Pastor-Satorras}}\ and\ \bibinfo {author} {\bibfnamefont {A.}~\bibnamefont
  {Vespignani}},\ }\href@noop {} {\bibfield  {journal} {\bibinfo  {journal}
  {Physical Review Letters}\ }\textbf {\bibinfo {volume} {86}},\ \bibinfo
  {pages} {3200} (\bibinfo {year} {2001})}\BibitemShut {NoStop}%
\bibitem [{\citenamefont {Wang}\ \emph
  {et~al.}(2013{\natexlab{b}})\citenamefont {Wang}, \citenamefont {Li},
  \citenamefont {D'Agostino}, \citenamefont {Havlin}, \citenamefont {Stanley},\
  and\ \citenamefont {Van~Mieghem}}]{wang2013effect}%
  \BibitemOpen
  \bibfield  {author} {\bibinfo {author} {\bibfnamefont {H.}~\bibnamefont
  {Wang}}, \bibinfo {author} {\bibfnamefont {Q.}~\bibnamefont {Li}}, \bibinfo
  {author} {\bibfnamefont {G.}~\bibnamefont {D'Agostino}}, \bibinfo {author}
  {\bibfnamefont {S.}~\bibnamefont {Havlin}}, \bibinfo {author} {\bibfnamefont
  {H.~E.}\ \bibnamefont {Stanley}}, \ and\ \bibinfo {author} {\bibfnamefont
  {P.}~\bibnamefont {Van~Mieghem}},\ }\href@noop {} {\bibfield  {journal}
  {\bibinfo  {journal} {Physical Review E}\ }\textbf {\bibinfo {volume} {88}},\
  \bibinfo {pages} {022801} (\bibinfo {year} {2013}{\natexlab{b}})}\BibitemShut
  {NoStop}%
\bibitem [{\citenamefont {de~Arruda}\ \emph {et~al.}(2017)\citenamefont
  {de~Arruda}, \citenamefont {Cozzo}, \citenamefont {Peixoto}, \citenamefont
  {Rodrigues},\ and\ \citenamefont {Moreno}}]{de2017disease}%
  \BibitemOpen
  \bibfield  {author} {\bibinfo {author} {\bibfnamefont {G.~F.}\ \bibnamefont
  {de~Arruda}}, \bibinfo {author} {\bibfnamefont {E.}~\bibnamefont {Cozzo}},
  \bibinfo {author} {\bibfnamefont {T.~P.}\ \bibnamefont {Peixoto}}, \bibinfo
  {author} {\bibfnamefont {F.~A.}\ \bibnamefont {Rodrigues}}, \ and\ \bibinfo
  {author} {\bibfnamefont {Y.}~\bibnamefont {Moreno}},\ }\href@noop {}
  {\bibfield  {journal} {\bibinfo  {journal} {Physical Review X}\ }\textbf
  {\bibinfo {volume} {7}},\ \bibinfo {pages} {011014} (\bibinfo {year}
  {2017})}\BibitemShut {NoStop}%
\bibitem [{\citenamefont {Dickison}\ \emph {et~al.}(2012)\citenamefont
  {Dickison}, \citenamefont {Havlin},\ and\ \citenamefont
  {Stanley}}]{dickison2012epidemics}%
  \BibitemOpen
  \bibfield  {author} {\bibinfo {author} {\bibfnamefont {M.}~\bibnamefont
  {Dickison}}, \bibinfo {author} {\bibfnamefont {S.}~\bibnamefont {Havlin}}, \
  and\ \bibinfo {author} {\bibfnamefont {H.~E.}\ \bibnamefont {Stanley}},\
  }\href@noop {} {\bibfield  {journal} {\bibinfo  {journal} {Physical Review
  E}\ }\textbf {\bibinfo {volume} {85}},\ \bibinfo {pages} {066109} (\bibinfo
  {year} {2012})}\BibitemShut {NoStop}%
\bibitem [{\citenamefont {Wang}\ \emph {et~al.}(2014)\citenamefont {Wang},
  \citenamefont {Tang}, \citenamefont {Yang}, \citenamefont {Do}, \citenamefont
  {Lai},\ and\ \citenamefont {Lee}}]{wang2014asymmetrically}%
  \BibitemOpen
  \bibfield  {author} {\bibinfo {author} {\bibfnamefont {W.}~\bibnamefont
  {Wang}}, \bibinfo {author} {\bibfnamefont {M.}~\bibnamefont {Tang}}, \bibinfo
  {author} {\bibfnamefont {H.}~\bibnamefont {Yang}}, \bibinfo {author}
  {\bibfnamefont {Y.}~\bibnamefont {Do}}, \bibinfo {author} {\bibfnamefont
  {Y.-C.}\ \bibnamefont {Lai}}, \ and\ \bibinfo {author} {\bibfnamefont
  {G.}~\bibnamefont {Lee}},\ }\href@noop {} {\bibfield  {journal} {\bibinfo
  {journal} {Scientific Reports}\ }\textbf {\bibinfo {volume} {4}},\ \bibinfo
  {pages} {5097} (\bibinfo {year} {2014})}\BibitemShut {NoStop}%
\bibitem [{\citenamefont {Wang}\ \emph
  {et~al.}(2016{\natexlab{a}})\citenamefont {Wang}, \citenamefont {Liu},
  \citenamefont {Cai}, \citenamefont {Tang}, \citenamefont {Braunstein},\ and\
  \citenamefont {Stanley}}]{wang2016suppressing}%
  \BibitemOpen
  \bibfield  {author} {\bibinfo {author} {\bibfnamefont {W.}~\bibnamefont
  {Wang}}, \bibinfo {author} {\bibfnamefont {Q.-H.}\ \bibnamefont {Liu}},
  \bibinfo {author} {\bibfnamefont {S.-M.}\ \bibnamefont {Cai}}, \bibinfo
  {author} {\bibfnamefont {M.}~\bibnamefont {Tang}}, \bibinfo {author}
  {\bibfnamefont {L.~A.}\ \bibnamefont {Braunstein}}, \ and\ \bibinfo {author}
  {\bibfnamefont {H.~E.}\ \bibnamefont {Stanley}},\ }\href@noop {} {\bibfield
  {journal} {\bibinfo  {journal} {Scientific Reports}\ }\textbf {\bibinfo
  {volume} {6}},\ \bibinfo {pages} {29259} (\bibinfo {year}
  {2016}{\natexlab{a}})}\BibitemShut {NoStop}%
\bibitem [{\citenamefont {Liu}\ \emph {et~al.}(2016)\citenamefont {Liu},
  \citenamefont {Wang}, \citenamefont {Tang},\ and\ \citenamefont
  {Zhang}}]{liu2016impacts}%
  \BibitemOpen
  \bibfield  {author} {\bibinfo {author} {\bibfnamefont {Q.-H.}\ \bibnamefont
  {Liu}}, \bibinfo {author} {\bibfnamefont {W.}~\bibnamefont {Wang}}, \bibinfo
  {author} {\bibfnamefont {M.}~\bibnamefont {Tang}}, \ and\ \bibinfo {author}
  {\bibfnamefont {H.-F.}\ \bibnamefont {Zhang}},\ }\href@noop {} {\bibfield
  {journal} {\bibinfo  {journal} {Scientific Reports}\ }\textbf {\bibinfo
  {volume} {6}} (\bibinfo {year} {2016})}\BibitemShut {NoStop}%
\bibitem [{\citenamefont {Parshani}\ \emph {et~al.}(2011)\citenamefont
  {Parshani}, \citenamefont {Rozenblat}, \citenamefont {Ietri}, \citenamefont
  {Ducruet},\ and\ \citenamefont {Havlin}}]{parshani2011inter}%
  \BibitemOpen
  \bibfield  {author} {\bibinfo {author} {\bibfnamefont {R.}~\bibnamefont
  {Parshani}}, \bibinfo {author} {\bibfnamefont {C.}~\bibnamefont {Rozenblat}},
  \bibinfo {author} {\bibfnamefont {D.}~\bibnamefont {Ietri}}, \bibinfo
  {author} {\bibfnamefont {C.}~\bibnamefont {Ducruet}}, \ and\ \bibinfo
  {author} {\bibfnamefont {S.}~\bibnamefont {Havlin}},\ }\href@noop {}
  {\bibfield  {journal} {\bibinfo  {journal} {Europhysics Letters}\ }\textbf
  {\bibinfo {volume} {92}},\ \bibinfo {pages} {68002} (\bibinfo {year}
  {2011})}\BibitemShut {NoStop}%
\bibitem [{\citenamefont {Aguirre}\ \emph {et~al.}(2013)\citenamefont
  {Aguirre}, \citenamefont {Papo},\ and\ \citenamefont
  {Buld{\'u}}}]{aguirre2013successful}%
  \BibitemOpen
  \bibfield  {author} {\bibinfo {author} {\bibfnamefont {J.}~\bibnamefont
  {Aguirre}}, \bibinfo {author} {\bibfnamefont {D.}~\bibnamefont {Papo}}, \
  and\ \bibinfo {author} {\bibfnamefont {J.~M.}\ \bibnamefont {Buld{\'u}}},\
  }\href@noop {} {\bibfield  {journal} {\bibinfo  {journal} {Nature Physics}\
  }\textbf {\bibinfo {volume} {9}},\ \bibinfo {pages} {230} (\bibinfo {year}
  {2013})}\BibitemShut {NoStop}%
\bibitem [{\citenamefont {Aguirre}\ \emph {et~al.}(2014)\citenamefont
  {Aguirre}, \citenamefont {Sevilla-Escoboza}, \citenamefont {Guti{\'e}rrez},
  \citenamefont {Papo},\ and\ \citenamefont
  {Buld{\'u}}}]{aguirre2014synchronization}%
  \BibitemOpen
  \bibfield  {author} {\bibinfo {author} {\bibfnamefont {J.}~\bibnamefont
  {Aguirre}}, \bibinfo {author} {\bibfnamefont {R.}~\bibnamefont
  {Sevilla-Escoboza}}, \bibinfo {author} {\bibfnamefont {R.}~\bibnamefont
  {Guti{\'e}rrez}}, \bibinfo {author} {\bibfnamefont {D.}~\bibnamefont {Papo}},
  \ and\ \bibinfo {author} {\bibfnamefont {J.}~\bibnamefont {Buld{\'u}}},\
  }\href@noop {} {\bibfield  {journal} {\bibinfo  {journal} {Physical Review
  Letters}\ }\textbf {\bibinfo {volume} {112}},\ \bibinfo {pages} {248701}
  (\bibinfo {year} {2014})}\BibitemShut {NoStop}%
\bibitem [{\citenamefont {Li}\ \emph {et~al.}(2016)\citenamefont {Li},
  \citenamefont {Wu}, \citenamefont {Lu},\ and\ \citenamefont
  {L{\"u}}}]{li2016synchronizability}%
  \BibitemOpen
  \bibfield  {author} {\bibinfo {author} {\bibfnamefont {Y.}~\bibnamefont
  {Li}}, \bibinfo {author} {\bibfnamefont {X.}~\bibnamefont {Wu}}, \bibinfo
  {author} {\bibfnamefont {J.-a.}\ \bibnamefont {Lu}}, \ and\ \bibinfo {author}
  {\bibfnamefont {J.}~\bibnamefont {L{\"u}}},\ }\href@noop {} {\bibfield
  {journal} {\bibinfo  {journal} {IEEE Transactions on Circuits and Systems II:
  Express Briefs}\ }\textbf {\bibinfo {volume} {63}},\ \bibinfo {pages} {206}
  (\bibinfo {year} {2016})}\BibitemShut {NoStop}%
\bibitem [{\citenamefont {Wei}\ \emph {et~al.}(2018{\natexlab{a}})\citenamefont
  {Wei}, \citenamefont {Wu}, \citenamefont {Lu},\ and\ \citenamefont
  {Wei}}]{wei2018synchronizability}%
  \BibitemOpen
  \bibfield  {author} {\bibinfo {author} {\bibfnamefont {J.}~\bibnamefont
  {Wei}}, \bibinfo {author} {\bibfnamefont {X.}~\bibnamefont {Wu}}, \bibinfo
  {author} {\bibfnamefont {J.-A.}\ \bibnamefont {Lu}}, \ and\ \bibinfo {author}
  {\bibfnamefont {X.}~\bibnamefont {Wei}},\ }\href@noop {} {\bibfield
  {journal} {\bibinfo  {journal} {Europhysics Letters}\ }\textbf {\bibinfo
  {volume} {120}},\ \bibinfo {pages} {20005} (\bibinfo {year}
  {2018}{\natexlab{a}})}\BibitemShut {NoStop}%
\bibitem [{\citenamefont {Wei}\ \emph {et~al.}(2018{\natexlab{b}})\citenamefont
  {Wei}, \citenamefont {Emenheiser}, \citenamefont {Wu}, \citenamefont {Lu},\
  and\ \citenamefont {D'Souza}}]{wei2018maximizing}%
  \BibitemOpen
  \bibfield  {author} {\bibinfo {author} {\bibfnamefont {X.}~\bibnamefont
  {Wei}}, \bibinfo {author} {\bibfnamefont {J.}~\bibnamefont {Emenheiser}},
  \bibinfo {author} {\bibfnamefont {X.}~\bibnamefont {Wu}}, \bibinfo {author}
  {\bibfnamefont {J.-A.}\ \bibnamefont {Lu}}, \ and\ \bibinfo {author}
  {\bibfnamefont {R.~M.}\ \bibnamefont {D'Souza}},\ }\href@noop {} {\bibfield
  {journal} {\bibinfo  {journal} {Chaos }\ }\textbf {\bibinfo {volume} {28}},\ \bibinfo {pages}
  {013110} (\bibinfo {year} {2018}{\natexlab{b}})}\BibitemShut {NoStop}%
\bibitem [{\citenamefont {Goltsev}\ \emph {et~al.}(2012)\citenamefont
  {Goltsev}, \citenamefont {Dorogovtsev}, \citenamefont {Oliveira},\ and\
  \citenamefont {Mendes}}]{Goltsev2012}%
  \BibitemOpen
  \bibfield  {author} {\bibinfo {author} {\bibfnamefont {A.}~\bibnamefont
  {Goltsev}}, \bibinfo {author} {\bibfnamefont {S.}~\bibnamefont
  {Dorogovtsev}}, \bibinfo {author} {\bibfnamefont {J.}~\bibnamefont
  {Oliveira}}, \ and\ \bibinfo {author} {\bibfnamefont {J.}~\bibnamefont
  {Mendes}},\ }\href@noop {} {\bibfield  {journal} {\bibinfo  {journal} {Physical\
  Review\ Letters}\ }\textbf {\bibinfo {volume} {109}},\ \bibinfo {pages} {128702}
  (\bibinfo {year} {2012})}\BibitemShut {NoStop}%
\bibitem [{\citenamefont {G{\'o}mez}\ \emph {et~al.}(2010)\citenamefont
  {G{\'o}mez}, \citenamefont {Arenas}, \citenamefont {Borge-Holthoefer},
  \citenamefont {Meloni},\ and\ \citenamefont {Moreno}}]{gomez2010discrete}%
  \BibitemOpen
  \bibfield  {author} {\bibinfo {author} {\bibfnamefont {S.}~\bibnamefont
  {G{\'o}mez}}, \bibinfo {author} {\bibfnamefont {A.}~\bibnamefont {Arenas}},
  \bibinfo {author} {\bibfnamefont {J.}~\bibnamefont {Borge-Holthoefer}},
  \bibinfo {author} {\bibfnamefont {S.}~\bibnamefont {Meloni}}, \ and\ \bibinfo
  {author} {\bibfnamefont {Y.}~\bibnamefont {Moreno}},\ }\href@noop {}
  {\bibfield  {journal} {\bibinfo  {journal} {Europhysics Letters}\ }\textbf
  {\bibinfo {volume} {89}},\ \bibinfo {pages} {38009} (\bibinfo {year}
  {2010})}\BibitemShut {NoStop}%
\bibitem [{\citenamefont {Goh}\ \emph {et~al.}(2001)\citenamefont {Goh},
  \citenamefont {Kahng},\ and\ \citenamefont {Kim}}]{Goh2001}%
  \BibitemOpen
  \bibfield  {author} {\bibinfo {author} {\bibfnamefont {K.~I.}\ \bibnamefont
  {Goh}}, \bibinfo {author} {\bibfnamefont {B.}~\bibnamefont {Kahng}}, \ and\
  \bibinfo {author} {\bibfnamefont {D.}~\bibnamefont {Kim}},\ }\href@noop {}
  {\bibfield  {journal} {\bibinfo  {journal} {Physical\ Review\ E}\ }\textbf
  {\bibinfo {volume} {64}},\ \bibinfo {pages} {051903} (\bibinfo {year}
  {2001})}\BibitemShut {NoStop}%
\bibitem [{\citenamefont {Massa}\ \emph {et~al.}(2009)\citenamefont {Massa},
  \citenamefont {Salvetti},\ and\ \citenamefont {Tomasoni}}]{Massa2009}%
  \BibitemOpen
  \bibfield  {author} {\bibinfo {author} {\bibfnamefont {P.}~\bibnamefont
  {Massa}}, \bibinfo {author} {\bibfnamefont {M.}~\bibnamefont {Salvetti}}, \
  and\ \bibinfo {author} {\bibfnamefont {D.}~\bibnamefont {Tomasoni}},\ }in\
  \href@noop {} {\emph {\bibinfo {booktitle} {Dependable, Autonomic and Secure
  Computing, 2009. DASC'09. Eighth IEEE International Conference on}}}\
  (\bibinfo {organization} {IEEE},\ \bibinfo {year} {2009})\ pp.\ \bibinfo
  {pages} {658--663}\BibitemShut {NoStop}%
\bibitem [{\citenamefont {Leskovec}\ and\ \citenamefont
  {Mcauley}(2012)}]{Leskovec2012}%
  \BibitemOpen
  \bibfield  {author} {\bibinfo {author} {\bibfnamefont {J.}~\bibnamefont
  {Leskovec}}\ and\ \bibinfo {author} {\bibfnamefont {J.~J.}\ \bibnamefont
  {Mcauley}},\ }in\ \href@noop {} {\emph {\bibinfo {booktitle} {Advances in
  neural information processing systems}}}\ (\bibinfo {year} {2012})\ pp.\
  \bibinfo {pages} {539--547}\BibitemShut {NoStop}%
\bibitem [{\citenamefont {Leskovec}\ \emph {et~al.}(2007)\citenamefont
  {Leskovec}, \citenamefont {Kleinberg},\ and\ \citenamefont
  {Faloutsos}}]{Leskovec2007}%
  \BibitemOpen
  \bibfield  {author} {\bibinfo {author} {\bibfnamefont {J.}~\bibnamefont
  {Leskovec}}, \bibinfo {author} {\bibfnamefont {J.}~\bibnamefont {Kleinberg}},
  \ and\ \bibinfo {author} {\bibfnamefont {C.}~\bibnamefont {Faloutsos}},\
  }\href@noop {} {\bibfield  {journal} {\bibinfo  {journal} {ACM Transactions
  on Knowledge Discovery from Data (TKDD)}\ }\textbf {\bibinfo {volume} {1}},\
  \bibinfo {pages} {2} (\bibinfo {year} {2007})}\BibitemShut {NoStop}%
\bibitem [{Kon()}]{Konect}%
  \BibitemOpen
  \href@noop {} {}\bibinfo {howpublished}
  {\url{http://konect.uni-koblenz.de/networks/}}\BibitemShut {NoStop}%
\bibitem [{\citenamefont {Newman}(2006)}]{newman2006modularity}%
  \BibitemOpen
  \bibfield  {author} {\bibinfo {author} {\bibfnamefont {M.~E.}\ \bibnamefont
  {Newman}},\ }\href@noop {} {\bibfield  {journal} {\bibinfo  {journal}
  {Proceedings of the National Academy of Sciences}\ }\textbf {\bibinfo
  {volume} {103}},\ \bibinfo {pages} {8577} (\bibinfo {year}
  {2006})}\BibitemShut {NoStop}%
\bibitem [{\citenamefont {Shu}\ \emph {et~al.}(2015)\citenamefont {Shu},
  \citenamefont {Wang}, \citenamefont {Tang},\ and\ \citenamefont
  {Do}}]{Shu2015}%
  \BibitemOpen
  \bibfield  {author} {\bibinfo {author} {\bibfnamefont {P.}~\bibnamefont
  {Shu}}, \bibinfo {author} {\bibfnamefont {W.}~\bibnamefont {Wang}}, \bibinfo
  {author} {\bibfnamefont {M.}~\bibnamefont {Tang}}, \ and\ \bibinfo {author}
  {\bibfnamefont {Y.}~\bibnamefont {Do}},\ }\href@noop {} {\bibfield  {journal}
  {\bibinfo  {journal} {Chaos}\ }\textbf {\bibinfo {volume} {25}},\ \bibinfo
  {pages} {063104} (\bibinfo {year} {2015})}\BibitemShut {NoStop}%
\bibitem [{\citenamefont {Shu}\ \emph {et~al.}(2016)\citenamefont {Shu},
  \citenamefont {Wang}, \citenamefont {Tang},\ and\ \citenamefont
  {Zhao}}]{Shu2016}%
  \BibitemOpen
  \bibfield  {author} {\bibinfo {author} {\bibfnamefont {P.}~\bibnamefont
  {Shu}}, \bibinfo {author} {\bibfnamefont {W.}~\bibnamefont {Wang}}, \bibinfo
  {author} {\bibfnamefont {M.}~\bibnamefont {Tang}}, \ and\ \bibinfo {author}
  {\bibfnamefont {P.}~\bibnamefont {Zhao}},\ }\href@noop {} {\bibfield
  {journal} {\bibinfo  {journal} {Chaos}\ }\textbf {\bibinfo {volume} {26}},\
  \bibinfo {pages} {063108} (\bibinfo {year} {2016})}\BibitemShut {NoStop}%
\bibitem [{\citenamefont {Wang}\ \emph {et~al.}(2017)\citenamefont {Wang},
  \citenamefont {Tang}, \citenamefont {Stanley},\ and\ \citenamefont
  {Braunstein}}]{wang2017unification}%
  \BibitemOpen
  \bibfield  {author} {\bibinfo {author} {\bibfnamefont {W.}~\bibnamefont
  {Wang}}, \bibinfo {author} {\bibfnamefont {M.}~\bibnamefont {Tang}}, \bibinfo
  {author} {\bibfnamefont {H.~E.}\ \bibnamefont {Stanley}}, \ and\ \bibinfo
  {author} {\bibfnamefont {L.~A.}\ \bibnamefont {Braunstein}},\ }\href@noop {}
  {\bibfield  {journal} {\bibinfo  {journal} {Reports on Progress in Physics}\
  }\textbf {\bibinfo {volume} {80}},\ \bibinfo {pages} {036603} (\bibinfo
  {year} {2017})}\BibitemShut {NoStop}%
\bibitem [{\citenamefont {Radicchi}\ and\ \citenamefont
  {Castellano}(2016)}]{radicchi2016beyond}%
  \BibitemOpen
  \bibfield  {author} {\bibinfo {author} {\bibfnamefont {F.}~\bibnamefont
  {Radicchi}}\ and\ \bibinfo {author} {\bibfnamefont {C.}~\bibnamefont
  {Castellano}},\ }\href@noop {} {\bibfield  {journal} {\bibinfo  {journal}
  {Physical Review E}\ }\textbf {\bibinfo {volume} {93}},\ \bibinfo {pages}
  {030302} (\bibinfo {year} {2016})}\BibitemShut {NoStop}%
\bibitem [{\citenamefont {Wang}\ \emph
  {et~al.}(2016{\natexlab{b}})\citenamefont {Wang}, \citenamefont {Liu},
  \citenamefont {Zhong}, \citenamefont {Tang}, \citenamefont {Gao},\ and\
  \citenamefont {Stanley}}]{wang2016predicting}%
  \BibitemOpen
  \bibfield  {author} {\bibinfo {author} {\bibfnamefont {W.}~\bibnamefont
  {Wang}}, \bibinfo {author} {\bibfnamefont {Q.-H.}\ \bibnamefont {Liu}},
  \bibinfo {author} {\bibfnamefont {L.-F.}\ \bibnamefont {Zhong}}, \bibinfo
  {author} {\bibfnamefont {M.}~\bibnamefont {Tang}}, \bibinfo {author}
  {\bibfnamefont {H.}~\bibnamefont {Gao}}, \ and\ \bibinfo {author}
  {\bibfnamefont {H.~E.}\ \bibnamefont {Stanley}},\ }\href@noop {} {\bibfield
  {journal} {\bibinfo  {journal} {Scientific Reports}\ }\textbf {\bibinfo
  {volume} {6}},\ \bibinfo {pages} {24676} (\bibinfo {year}
  {2016}{\natexlab{b}})}\BibitemShut {NoStop}%
\bibitem [{\citenamefont {Gleeson}\ \emph {et~al.}(2012)\citenamefont
  {Gleeson}, \citenamefont {Melnik}, \citenamefont {Ward}, \citenamefont
  {Porter},\ and\ \citenamefont {Mucha}}]{gleeson2012accuracy}%
  \BibitemOpen
  \bibfield  {author} {\bibinfo {author} {\bibfnamefont {J.~P.}\ \bibnamefont
  {Gleeson}}, \bibinfo {author} {\bibfnamefont {S.}~\bibnamefont {Melnik}},
  \bibinfo {author} {\bibfnamefont {J.~A.}\ \bibnamefont {Ward}}, \bibinfo
  {author} {\bibfnamefont {M.~A.}\ \bibnamefont {Porter}}, \ and\ \bibinfo
  {author} {\bibfnamefont {P.~J.}\ \bibnamefont {Mucha}},\ }\href@noop {}
  {\bibfield  {journal} {\bibinfo  {journal} {Physical Review E}\ }\textbf
  {\bibinfo {volume} {85}},\ \bibinfo {pages} {026106} (\bibinfo {year}
  {2012})}\BibitemShut {NoStop}%
\bibitem [{\citenamefont {L{\"u}}\ \emph {et~al.}(2016)\citenamefont {L{\"u}},
  \citenamefont {Chen}, \citenamefont {Ren}, \citenamefont {Zhang},
  \citenamefont {Zhang},\ and\ \citenamefont {Zhou}}]{lu2016vital}%
  \BibitemOpen
  \bibfield  {author} {\bibinfo {author} {\bibfnamefont {L.}~\bibnamefont
  {L{\"u}}}, \bibinfo {author} {\bibfnamefont {D.}~\bibnamefont {Chen}},
  \bibinfo {author} {\bibfnamefont {X.-L.}\ \bibnamefont {Ren}}, \bibinfo
  {author} {\bibfnamefont {Q.-M.}\ \bibnamefont {Zhang}}, \bibinfo {author}
  {\bibfnamefont {Y.-C.}\ \bibnamefont {Zhang}}, \ and\ \bibinfo {author}
  {\bibfnamefont {T.}~\bibnamefont {Zhou}},\ }\href@noop {} {\bibfield
  {journal} {\bibinfo  {journal} {Physics Reports}\ }\textbf {\bibinfo {volume}
  {650}},\ \bibinfo {pages} {1} (\bibinfo {year} {2016})}\BibitemShut {NoStop}%
\end{thebibliography}

\providecommand{\noopsort}[1]{}\providecommand{\singleletter}[1]{#1}%

\end{document}